\documentclass[sigconf]{acmart}


\usepackage{subcaption}
\usepackage{soul}
\usepackage{xspace}
\usepackage{multirow}
\usepackage{xurl}
\usepackage{url}
\usepackage[many]{tcolorbox}
\usepackage{enumitem}
\usepackage{hyperref}
\usepackage{booktabs}
\usepackage{pifont}
\usepackage{float}
\usepackage[framemethod=TikZ]{mdframed}
\usepackage{tikz}
\usepackage{graphicx}
\usepackage{stackengine}
\usepackage{makecell}
\definecolor{deepseek-r}{RGB}{91, 181, 172}    % Deepseek with reasoning (r1)
\definecolor{deepseek-nr}{RGB}{139, 218, 210}   % Deepseek without reasoning (v3)
\definecolor{qwen-r}{RGB}{216, 179, 101}        % Qwen with reasoning
\definecolor{qwen-nr}{RGB}{235, 203, 136}       % Qwen without reasoning
\definecolor{llama-r}{RGB}{222, 82, 108}        % LLaMA with reasoning
\definecolor{llama-nr}{RGB}{243, 126, 148}      % LLaMA without reasoning

\definecolor{gpt-r}{RGB}{130, 176, 210}

\definecolor{darkgrey}{HTML}{434343}
\definecolor{darkred}{HTML}{800000}
\definecolor{darkgreen}{HTML}{006400}

\newcommand{\framework}{\textsf{CORRECT}\xspace}

\newcommand{\halfcheckcross}{\stackinset{c}{-0.1ex}{c}{0ex}{\ding{51}}{\ding{55}}}
\newcommand{\blackcheck}{\ding{51}}
\newcommand{\blackcross}{\ding{55}}

\definecolor{darkgreen}{RGB}{0, 100, 0}   % Dark green
\definecolor{darkred}{RGB}{139, 0, 0}     % Dark red
\definecolor{darkyellow}{RGB}{139, 139, 0} % Dark yellow

\newcommand{\greencheck}{\textcolor{deepseek-r}{\blackcheck}}         % Green checkmark
\newcommand{\redcross}{\textcolor{llama-r}{\blackcross}}                 % Red cross
\newcommand{\yellowhalfcheck}{\textcolor{qwen-r}{\halfcheckcross}}

\newtcolorbox{mybox}[2][]{text width=0.95\linewidth,fontupper=\normalsize,
fonttitle=\bfseries\sffamily\scriptsize, colbacktitle=darkgrey,enhanced,
attach boxed title to top left={yshift=-2mm,xshift=3mm},
boxed title style={sharp corners},top=4pt,bottom=2pt,left=2pt,right=2pt,
  title=#2,colback=white}

\newcommand{\myframed}[2]{%
  \begin{mdframed}[linecolor=#1!90,backgroundcolor=#1!4,roundcorner=2pt,linewidth=1.5pt]
  #2
  \end{mdframed}
}

\newcounter{misunderstandingcounter}
\newcounter{findingcounter}
\newcounter{suggestioncounter}

\newcommand{\misbox}[2]{%
  \refstepcounter{misunderstandingcounter} % 允许引用
  \label{mis:#1} % 用 name 作为 label
  \myframed{llama-nr}{\textbf{Consensus \#\themisunderstandingcounter - #1:} #2 \looseness=-1} % 自动编号 + name + 内容
}

\newcommand{\misref}[1]{Consensus \#\ref{mis:#1}}

\newcommand{\findbox}[1]{%
  \stepcounter{findingcounter} % 增加 findbox 的计数器
  \myframed{qwen-nr}{\textbf{Finding \#\thefindingcounter: }#1} % 自动编号
}

\begin{document}

\title{Everything You Wanted to Know About LLM-based Vulnerability Detection But Were Afraid to Ask}
% \title{\framework: Context-Rich Reasoning Evaluation of LLM-based Code Vulnerability Detection} \hao{tentative, to discuss}

\author{Yue Li}
\affiliation{%
  \institution{National Key Lab for Novel Software
\\ Technology, Nanjing University}
  \city{Nanjing, Jiangsu}
  \country{China}}
\author{Xiao Li}
\affiliation{%
  \institution{National Key Lab for Novel Software
\\ Technology, Nanjing University}
  \city{Nanjing, Jiangsu}
  \country{China}}
  \author{Hao Wu}
\affiliation{%
  \institution{National Key Lab for Novel Software
\\ Technology, Nanjing University}
  \city{Nanjing, Jiangsu}
  \country{China}}
  \author{Minghui Xu}
\affiliation{%
  \institution{Shandong University}
  \city{Qingdao, Shandong}
  \country{China}}
  \author{Yue Zhang}
\affiliation{%
  \institution{Shandong University}
  \city{Qingdao, Shandong}
  \country{China}}
    \author{Xiuzhen Cheng}
\affiliation{%
  \institution{Shandong University}
  \city{Qingdao, Shandong}
  \country{China}}
    \author{Fengyuan Xu}
\affiliation{%
  \institution{National Key Lab for Novel Software
\\ Technology, Nanjing University}
  \city{Nanjing, Jiangsu}
  \country{China}}
    \author{Sheng Zhong}
\affiliation{%
  \institution{National Key Lab for Novel Software
\\ Technology, Nanjing University}
  \city{Nanjing, Jiangsu}
  \country{China}}

\begin{abstract}

Large Language Models have emerged as a promising tool for automated vulnerability detection, driven by their success in code generation, repair, and integration into developer tools.   However, despite widespread adoption, a critical question remains unresolved: \textit{Are LLMs truly effective at detecting real-world vulnerabilities?} Current evaluations, which often assess models on isolated functions or files, ignore the broader execution and data-flow context essential for understanding real-world vulnerabilities. This oversight leads to two types of misleading outcomes: incorrect conclusions and flawed rationales, collectively undermining the reliability of prior assessments.
Therefore, in this paper,  we challenge three widely held community beliefs: that LLMs are (i) unreliable, (ii) insensitive to code patches, and (iii) performance-plateaued across model scales. We argue that these beliefs are artifacts of context-deprived evaluations. To address this, we propose \framework (\textit{Context-Rich Reasoning Evaluation of Code with Trust}), a new evaluation framework that systematically incorporates contextual information into LLM-based vulnerability detection. We construct a context-rich dataset of 2,000 vulnerable–patched program pairs spanning 99 CWEs and evaluate 13 LLMs across four model families. Our framework elicits both binary predictions and natural-language rationales, which are further validated using LLM-as-a-judge techniques.
Our findings overturn existing misconceptions. When provided with sufficient context, state-of-the-art LLMs achieve significantly improved performance (e.g., 67\% accuracy and >70\% F1-score on key CWEs), with precision nearing 0.8. We show that most false positives stem from reasoning errors rather than misclassification, and that while model and test-time scaling improve performance, they introduce diminishing returns and trade-offs in recall. Finally, we uncover new flaws in current LLM-based detection systems, such as limited generalization and overthinking biases.

\end{abstract}


\maketitle
\section{Introduction}

In recent years, Large Language Models (LLMs) have rapidly gained traction in the field of vulnerability detection. Their strong performance in a variety of code-related tasks, ranging from code generation to program repair, has led to widespread adoption in security-critical workflows. Industry reports and academic trends both reflect this momentum: LLMs were embedded into developer tools such as GitHub Copilot~\cite{github_copilot} and Cursor~\cite{cursor_ai_editor}, offering inline suggestions and real-time vulnerability analysis. On the research front, a growing number of studies~\cite{ullah2024llms, ding2024vulnerability, ma2024combining, sheng2025large} have evaluated LLMs for detecting logic bugs and security vulnerabilities, suggesting a promising future for LLM-driven secure coding. This enthusiasm is further fueled by a practical need: with over 20,000 CVEs reported in 2024~\cite{cve2025}, organizations are under increasing pressure to reduce detection costs, speed up mitigation, and handle complex, cross-language codebases~\cite{cost_of_data_breach_2024_ibm}—all areas where LLMs seem to excel. \looseness=-1

Yet, despite this growing attention, a fundamental question remains unanswered: \textit{Are LLMs actually good at detecting real-world vulnerabilities?} While recent work has attempted to answer this by benchmarking LLMs on isolated code functions or files~\cite{ding2024vulnerability, ullah2024llms, yin2024multitask, steenhoek2024err}, these evaluations often ignore the broader execution or data-flow context that governs real-world vulnerabilities. In real software systems, critical logic is distributed across multiple functions and files, meaning that vulnerabilities may only become apparent when viewed in context. This gap between realistic scenarios and the granularity of current evaluations has led us to a key concern: current benchmarks may not be measuring what truly matters.
Specifically, we find that context-free evaluations often result in two types of misleading outcomes: \textbf{UO(I)} Incorrect Conclusion: The LLM reaches a plausible but incorrect decision (e.g., falsely reporting a patched program as vulnerable) due to insufficient input, misaligning with the ground-truth label. \textbf{UO(II)} Incorrect Rationale: The LLM may guess the label correctly but for the wrong reason, missing the actual vulnerability while citing irrelevant code fragments. These outcomes suggest that the reliability of past LLM evaluations has been significantly compromised by contextual blindness. 

Nonetheless, the security community has already formed a set of influential (but potentially flawed) consensus beliefs based on various evaluations (none of which have accounted for context):  

\begin{itemize} [partopsep=2pt, topsep=-\parskip, parsep=2pt, itemsep=2pt, leftmargin=*]
    \item \textbf{Consensus \#1- Unreliable.} LLMs' ability to detect vulnerabilities is comparable to or even worse than random guessing, suggesting that they struggle to effectively identify vulnerabilities.
    \item \textbf{Consensus \#2- Insensitive.}  In pair-wise evaluations, LLMs exhibit precision levels that are only slightly better than random guessing, coupled with a significantly high (1, 1) proportion, meaning that LLMs treat both the original vulnerable code and the patched code as having nearly equal likelihoods of being vulnerable. This highlights their inability to reliably differentiate patched code from its original vulnerable counterpart.
    \item \textbf{Consensus \#3- Plateaued.} The vulnerability detection performance of LLMs shows only minor differences across various architectures and sizes. Furthermore, as models become more capable or increase in scale, their ability to detect vulnerabilities does not show meaningful improvement.
\end{itemize}

This brings us to a critical turning point: Are these consensus beliefs valid, or are they artifacts of flawed evaluation pipelines? To answer this, we propose \framework (\textit{Context-Rich Reasoning Evaluation of Code with Trust}), a new framework for systematically evaluating LLMs in vulnerability detection under context-rich settings.  Our core idea is to comprehensively collect as much relevant contextual information as possible—such as callee functions and type declarations—to support the LLM during the detection process, while leveraging vulnerability information to focus solely on the ground-truth vulnerabilities with proper context during evaluation. \framework proceeds in three stages. First, we construct a new dataset containing 2,000 vulnerable–patched code pairs across 99 CWEs, each embedded with relevant execution and data context to support accurate detection. Second, we design context-rich prompts to elicit both binary predictions and human-readable rationales from the LLMs. Finally, we employ an LLM-as-a-judge mechanism to verify that the rationales correctly identify the root cause of each vulnerability and avoid misclassifying patched vulnerabilities as active. This design allows us to assess not just whether a model gets the right answer, but \textit{why}. \looseness=-1

Our evaluation spans 13 LLMs across four major model families (Qwen, Llama, DeepSeek, and OpenAI), covering sizes from 7B to 671B. Our results provide strong evidence that the three community-wide consensuses are not just inaccurate—they are misunderstandings caused by the absence of context in prior benchmarks. We summarize our key findings as follows:

\begin{itemize} [partopsep=2pt, topsep=-\parskip, parsep=2pt, itemsep=2pt, leftmargin=*]
    \item  \textbf{Contextual Information Reveals Underestimated Model Performance.}
Prior evaluations that omitted contextual information significantly underestimated the true capability of LLMs in vulnerability detection. Under a context-rich evaluation framework, the state-of-the-art model (DeepSeek-R1) achieves an accuracy of 67\%, with particularly strong performance on structurally consistent vulnerability categories such as CWE-664 (e.g. out-of-bounds access) and CWE-682 (e.g. integer overflow), where the F1-score exceeds 70\%. These results demonstrate that, when supplied with appropriate context, LLMs can reason effectively about a wide range of vulnerability types.
    \item  \textbf{Discrimination Is Not the Bottleneck, but Reasoning Limitations Obscure True Capability.}
LLMs demonstrate a strong ability to differentiate between vulnerable and patched code, achieving precision scores approaching 0.8 and exhibiting a relatively low rate of misclassifying patched code as vulnerable (\textasciitilde 10\%). Detailed analysis reveals that only a small subset of false positives (\textasciitilde 9.5\%) result from an actual failure to recognize the presence of a patch. The majority arise from reasoning errors—specifically, the incorrect belief that the applied patch is incomplete. This indicates that the primary challenge lies not in binary classification, but in accurately identifying and reasoning about the root cause of vulnerabilities.

\item \textbf{Scaling Enhances Performance, but with Diminishing Returns and New Trade-offs.}
Both model scaling (i.e., increasing parameter size) and test-time scaling (i.e., extending reasoning steps or output length) contribute to improved detection performance. However, the observed improvements are often incremental and come at a cost. For example, increasing the number of reasoning tokens fivefold yields only a modest 5\% gain in accuracy. Moreover, extended reasoning introduces conservative biases that reduce recall by approximately 10\%, suggesting a fundamental trade-off between precision and vulnerability assessment. These findings highlight that while scaling is beneficial, it is not a panacea and must be approached with caution.

\end{itemize}

\noindent\textbf{Contributions:} This paper makes the following contributions:

\begin{itemize} [partopsep=2pt, topsep=-\parskip, parsep=2pt, itemsep=2pt, leftmargin=*]
    \item We propose \framework, a novel framework that addresses the evaluation flaws caused by missing context in vulnerability detection. \framework provides a high-quality dataset with 98\% label accuracy, containing 2,000 program pairs and covering 99 CWEs. To date, it is the most context-rich dataset available for vulnerability detection. Our dataset and code are available\footnote{\url{https://anonymous.4open.science/r/CORRECT}}.
 
    \item We demonstrate that the three prevailing consensuses within the community are, in fact, misconceptions, stemming from flawed evaluation methodologies that overlook contextual information and consequently lead to erroneous conclusions. By rectifying these misconceptions, we provide evidence that the vulnerability detection capabilities of LLMs have been underestimated.
    
    \item We identify new flaws in LLM-based vulnerability detection, including the inability to accurately detect vulnerabilities, a significant lack of generalization capability for out-of-distribution vulnerabilities, overthinking tendencies in reasoning models, and limitations of test-time scaling. These findings provide a clearer direction for future work.
\end{itemize}

\section{Background and Related Work}
% \liyue{reasoning for code}
\subsection{Reasoning with LLMs}
\label{subsec:backgroundreasoning}

Psychologists describe two different ways the human brain forms thoughts~\cite{sloman1996empirical, kahneman2011thinking}, i.e., \textbf{System 1} and \textbf{System 2}.
System 1 involves fast, intuitive, and unconscious processing and operates automatically and effortlessly, like stopping automatically when the traffic light turns red. 
In contrast, System 2, a.k.a., \textbf{reasoning}, is characterized by slow, deliberate, and logical processes that require conscious effort and careful thought, such as planning a multi-day trip itinerary.
Reasoning is regarded as a fundamental human ability extensively studied in psychology and cognitive science~\cite{stenning2012human}.

\begin{figure}[h]
    \centering
    \includegraphics[width=0.48\textwidth]{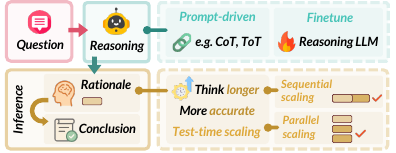}
    \vspace{-0.5cm}
    \caption{Reasoning with LLMs.}
    \label{fig:reasoning}
\end{figure}

Recent works~\cite{yu2024distilling, ji2025test} have shown that LLMs can perform both System 1 and System 2 thinking.  
System 1 thinking in LLMs refers to directly generating an answer without explicit processes.  
System 2 thinking, i.e., reasoning, involves the LLM engaging in deliberate thought before arriving at a final conclusion.  
The intermediate output of this reasoning process is called the \textbf{rationale}, while the final output is referred to as the \textbf{conclusion}, as shown in \autoref{fig:reasoning}.

Compared to System 1 thinking, reasoning achieves higher accuracy by decomposing complex tasks into sequential steps.
It also enhances interpretability, as the intermediate rationale provides transparent insights into LLM's decision-making process.
Recent works further demonstrate that LLMs exhibit \textit{testing-time scaling} properties~\cite{snell2024scaling}—allocating more computation time (e.g., additional reasoning steps) during inference improves LLMs' accuracy. \textit{Testing-time scaling} falls into two categories~\cite{muennighoff2025s1}: \textit{sequential scaling}, where later computations depend on earlier ones (e.g., extending a single rationale), and \textit{parallel scaling}, where computations run independently (e.g., majority voting).
Consequently, reasoning has been widely adopted across domains to enhance answer accuracy~\cite{hendrycks2021measuring, ahn2024large, ding2024reasoning, bhargava2022commonsense}. \looseness=-1

LLMs can perform reasoning in two distinct paradigms~\cite{huang2022towards}. 
The first, referred to as \textbf{prompt-driven reasoning}, leverages techniques such as Chain-of-Thought (CoT)~\cite{wei2022chain} and Tree-of-Thought (ToT)~\cite{yao2023tree} to enable a general LLM to engage in a ``thinking'' process before generating a response. 
The second, known as \textbf{fine-tuned reasoning}, involves using reinforcement learning~\cite{guo2025deepseek, havrilla2024teaching, lightman2023let} for reasoning tasks or fine-tuning the LLM on reasoning-specific datasets via supervised fine-tuning (SFT)~\cite{rajani2019explain, hendrycks2021measuring}.  
LLMs fine-tuned in this manner are also referred to as \textit{reasoning LLMs}, e.g., DeepSeek-R1~\cite{guo2025deepseek} and o3-mini~\cite{openai_o3_mini_system_card_2025}.  
In contrast, general LLMs that have not been fine-tuned specifically for reasoning tasks are referred to as \textit{non-reasoning LLMs}.

\subsection{LLM for Vulnerability Detection}
\label{subsec:llm-for-vd}
The remarkable performance of LLMs in code-related tasks has driven their rapid adoption for vulnerability detection. Initially, the community relied on basic queries, where the LLM was simply provided with the target code and asked to identify potential vulnerabilities~\cite{gao2023far}. While techniques like fine-tuning~\cite{ding2024vulnerability} can improve the accuracy of LLM-based vulnerability detection, they often require modifying model parameters and retraining the model, making them less practical for many applications. Consequently, more attention has been focused on approaches that enhance detection capabilities through prompt engineering or additional data, which do not require altering the core model. Subsequent improvements have adopted two main approaches to further boost detection accuracy:\looseness=-1

\begin{itemize}

\item \textbf{RAG-enhanced Vulnerability Detection:}
Retrieval Augmented Generation (RAG)~\cite{lewis2020retrieval, fan2024survey} enhances the accuracy of LLM responses by retrieving relevant knowledge from external sources and augmenting input prompts with this information. This approach addresses the limitations of LLMs in domain-specific knowledge gaps while mitigating hallucinations by grounding responses in retrieved evidence. In vulnerability detection, RAG retrieves code snippets with known vulnerabilities or patches that are semantically similar to the target code~\cite{du2024vul, steenhoek2024err}. However, it relies solely on semantic similarity to determine which external knowledge to incorporate without guaranteeing that the target code shares the same vulnerabilities as the retrieved snippets. A RAG example is presented in Appendix \ref{appendix:rag}.

\item \textbf{Reasoning-enhanced Vulnerability Detection:}
Recent studies show that reasoning significantly improves both accuracy and interpretability in LLM-based vulnerability detection systems~\cite{sun2024llm4vuln, ullah2024llms, nong2024chain}. 
Reasoning leverages LLMs' advanced logical capabilities by decomposing complex problems into smaller steps~\cite{nong2024chain}. 
It enables LLMs to identify subtle code flaws—such as intricate control flows or unsafe data handling—that non-reasoning methods often miss.
% logical issues in code, such as complex control flows or improper data handling - vulnerabilities that are both common and challenging to detect. 
For example, reasoning-enhanced methods can analyze whether a variable might become NULL under specific conditions, exposing potential Null Pointer Dereference vulnerabilities. Moreover, during the reasoning process, LLMs demonstrate self-reflection~\cite{pan2024automatically} and adaptive decision-making, allowing them to refine their analysis and correct errors. This iterative refinement mechanism effectively enhances vulnerability reasoning~\cite{ristea2024benchmarking}.
  
\end{itemize}
\begin{table}[t]
    \centering
    \caption{Evaluations of LLM-based vulnerability detection.}
    \vspace{-0.2cm}
    \label{tab:evaluation_comparison}
    \resizebox{\columnwidth}{!}{% Resize table to fit column width
    \footnotesize
    \begin{tabular}{>{}lcccc} % 第一列字体改为 \small
        \toprule
        \textbf{Work} & 
        \makecell{\textbf{Real} \\ \textbf{World} \\ \textbf{Dataset}} &
        \makecell{\textbf{Mitigating} \\ \textbf{Context} \\ \textbf{Limitation}} & 
        \makecell{\textbf{Label} \\ \textbf{Accuracy}} & 
        \makecell{\textbf{Rationale} \\ \textbf{Evaluation}} \\
        \midrule
        SecLLMHolmes~\cite{ullah2024llms} & \yellowhalfcheck & \redcross & \greencheck & \yellowhalfcheck \\
        \makecell[l]{
        Steenhoek et al.~\cite{steenhoek2024err}, PrimeVul~\cite{ding2024vulnerability}} & \greencheck & \redcross & \greencheck & \redcross \\
        \midrule
        \makecell[l]{VulEval~\cite{wen2024vuleval}\\ VulnSage~\cite{zibaeirad2025reasoning}, JitVul~\cite{yildiz2025benchmarking}
        } & 
        \greencheck & \yellowhalfcheck & \greencheck & \redcross \\
        \midrule
        \makecell[l]{VulnLLMEval \cite{zibaeirad2024vulnllmeval}, SCoPE \cite{gonccalves2024scope} \\ Yin et al. \cite{yin2024multitask}, Khare et al. \cite{khare2023understanding}} &
        \greencheck & \redcross & \redcross & \redcross \\
        \midrule
        VulDetectBench~\cite{liu2024vuldetectbench} & \yellowhalfcheck & \redcross & \redcross & \redcross \\
        \makecell[l]{Gao et al. \cite{gao2023far}, LLM4Vuln \cite{sun2024llm4vuln}} & 
        \yellowhalfcheck & \yellowhalfcheck & \redcross & \redcross \\
        \midrule
        \textbf{\framework} & \greencheck & \greencheck & \greencheck & \greencheck \\ % Added row
        \bottomrule
    \end{tabular}}

\end{table}

\vspace{2mm}\noindent\textbf{Limitations of Existing Evaluations}.
\autoref{tab:evaluation_comparison} summarizes existing evaluations of LLMs' vulnerability detection capabilities. These works suffer from three key limitations that lead to biased or unreliable conclusions.  {First, missing context.}
As illustrated in \S\ref{sec:motivation}, evaluations based on isolated functions or files (e.g., ~\cite{ullah2024llms, ding2024vulnerability, steenhoek2024err, liu2024vuldetectbench}) produce unreliable outcomes due to insufficient contextual information. Some works perform context augmentation by incorporating caller-callee relationships in the code to address the issues~\cite{wen2024vuleval, sun2024llm4vuln, gao2023far, yildiz2025benchmarking}. However, they fail to capture complex syntactic dependencies beyond direct function calls. 
To mitigate the impact of missing context, VulnSage~\cite{zibaeirad2025reasoning} introduces ``not sure'' responses as an alternative to binary classifications. However, it fails to fundamentally address the context-missing issue.
 {Second, low label accuracy.}
Work~\cite{croft2023data} demonstrates that flawed data collection methods frequently yield low label accuracy, severely skewing evaluation results. Multiple studies~\cite{yin2024multitask, zibaeirad2024vulnllmeval, gonccalves2024scope, sun2024llm4vuln, liu2024vuldetectbench, gao2023far, khare2023understanding} employ datasets with label accuracy as low as 20\%–71\%~\cite{croft2023data}, critically compromising their findings' validity. 
 {Third, overlooked reasoning.} Existing studies lack rationale evaluation, representing a critical disconnect from the widespread adoption of reasoning techniques in vulnerability detection. SecLLMHolmes~\cite{ullah2024llms} pioneers rationale assessment by comparing them against manual ground-truth annotations, but this approach is labor-intensive and non-scalable.  

\section{Motivation and Problem Statement}
\label{sec:motivation}

\begin{figure*}[t] 

    \centering
    \includegraphics[width=\textwidth]{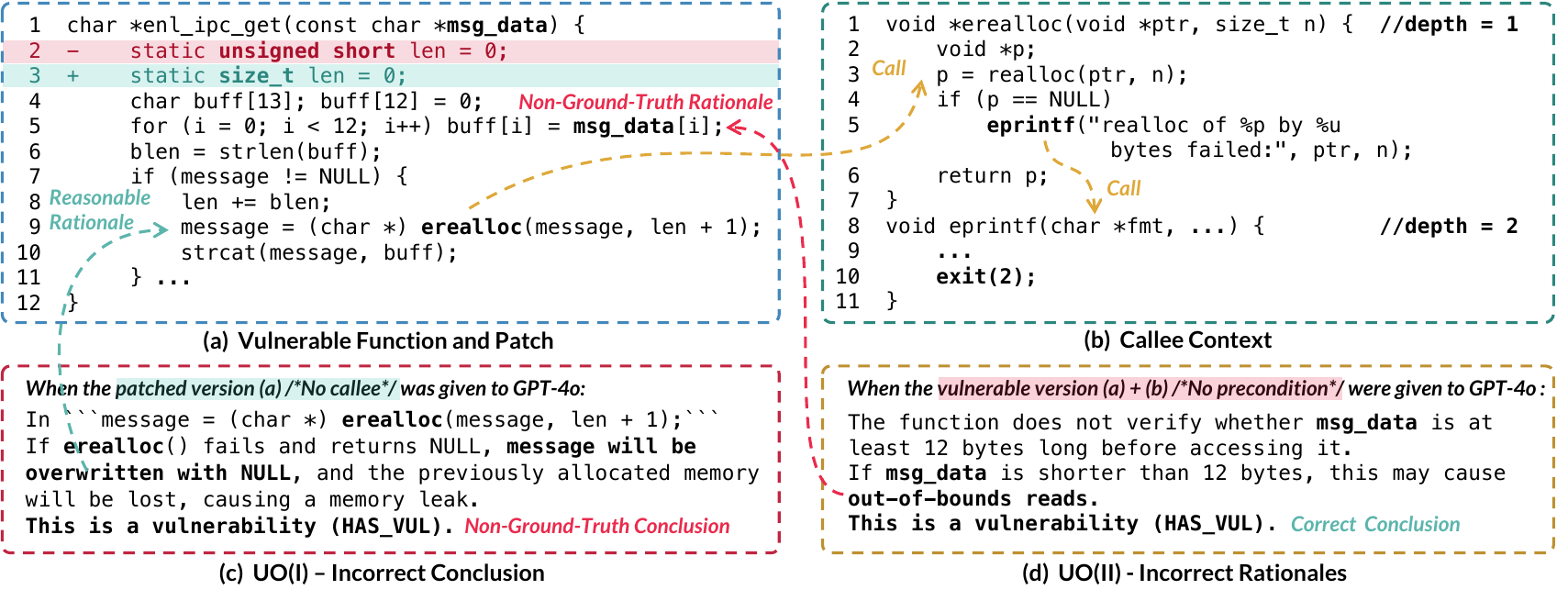} % 
    \vspace{-0.5cm}
    \caption{(a) CVE-2017-7875, an out-of-bounds write vulnerability. (b) Essential callee context for (a). (c) A reasonable rationale but an incorrect conclusion. (d) A correct conclusion, but a wrong rationale (non-ground-truth vulnerability).}
    \label{fig:code}
\end{figure*}

\subsection{Motivation}
\label{subsec:examples}

As the reasoning capabilities of LLMs continue to advance, prompt-driven reasoning methods (e.g., Chain-of-Thought~\cite{wei2022chain}) and reasoning LLMs (e.g., DeepSeek-R1~\cite{guo2025deepseek}, o3-mini~\cite{openai_o3_mini_system_card_2025}) are becoming increasingly mainstream. 
Those reasoning-centric LLMs not only predict the presence of vulnerabilities but also provide detailed rationales to justify their conclusions. However, most existing studies on LLM-based vulnerability detection remain limited to function-level or file-level analysis~\cite{sheng2025llms}, i.e., examining isolated snippets of code extracted from individual functions or files. This narrow scope often overlooks the broader program context that governs the code's behavior, which is critical for accurate reasoning. As a result, limited context can lead to undesired outcomes (\textbf{UO}) in vulnerability detection, affecting either the conclusions or the underlying rationales: \looseness=-1

\vspace{1mm}
\noindent\textbf{UO(I) -- Incorrect Conclusion:} Missing context can lead LLMs to produce logically sound explanations, while the actual conclusion (i.e., whether a vulnerability exists) may still be wrong. \autoref{fig:code} illustrates this issue using an example of an out-of-bounds write vulnerability. As shown in \autoref{fig:code}, the vulnerability originates \texttt{enl\_ipc\_get} function (\autoref{fig:code}(a)), where the \texttt{len} variable (declared as \texttt{unsigned short}) overflows when processing inputs larger than 64\,KB (line 8). This leads to incorrect memory allocation via \texttt{erealloc} (line 9), enabling an out-of-bounds write during \texttt{strcat} execution (line 10). The official patch replaces \texttt{len}'s type with \texttt{size\_t}, theoretically supporting up to 16 exabytes of data.  However, the patch's validity hinges on \texttt{erealloc}'s implementation context. Without \autoref{fig:code}(b), the patched code could still be deemed vulnerable: if \texttt{erealloc} fails, LLMs might assume standard \texttt{realloc} semantics (returning \texttt{NULL}), causing \texttt{strcat} to dereference a null pointer. In reality, \texttt{erealloc} terminates the program via \texttt{eprintf} (lines 5, 10 in \autoref{fig:code}(b)), preventing the vulnerability.

This contextual gap critically impacts evaluation accuracy, as shown in \autoref{fig:code}(c): When analyzing only \autoref{fig:code}(a), GPT-4o logically infers potential \texttt{erealloc} failures and flags \texttt{strcat} as vulnerable, outputting \texttt{HAS\_VUL}, which conflicts with the ground-truth label (patched code marked \texttt{NO\_VUL}). However, even when the predicted label is inconsistent with the ground-truth, the rationale is reasonable without context - even human experts cannot reliably judge vulnerabilities without full context, let alone AI models. This can lead to poor detection performance being incorrectly attributed to the model's deficiencies during evaluating its capabilities, when in fact, it is caused by the evaluation's shortcomings.

\vspace{1mm}
\noindent\textbf{UO(II) -- Incorrect Rationales:} 
Missing context can also lead an LLM to arrive at the correct predicted label based on an incorrect rationale. That is, when presented with a vulnerable function, an LLM may only correctly determine that a vulnerability exists. However, this does not necessarily mean that the model has accurately identified the true nature or location of the ground-truth vulnerability. Instead, the model might erroneously flag non-ground-truth sections of the code as vulnerable. For instance, consider the \textsf{msg\_data} parameter in \autoref{fig:code}(a). On line 5, there is an implicit assumption that \textsf{msg\_data} is at least 12 bytes in size, which is valid by default. However, since the LLM lacks this context (and it's almost impossible to provide this kind of contextual knowledge in practice), it concludes that line 5 contains a vulnerability (\autoref{fig:code}(d)) and predicts the label \texttt{HAS\_VUL}. However, even though the predicted label matches the ground-truth label \texttt{HAS\_VUL}, this does not mean that the LLM has truly detected the ground-truth vulnerability. Traditional evaluation methods based solely on predicted labels would mistakenly assume that the LLM produced the correct result, thereby overestimating the model's ability.

\subsection{SOTA Consensus and Problem Statement}
\label{subsec:rq}

From \S\ref{subsec:examples}, it is evident that the absence of context can lead to flawed reasoning and incorrect conclusions in vulnerability detection, thereby compromising the validity of evaluations. However, existing evaluation methodologies have not systematically examined the impact of missing context. This naturally raises concerns about the objectivity and reliability of conclusions drawn by current state-of-the-art (SOTA) approaches. In what follows, we present three widely recognized consensus points regarding the limitations of LLM-based vulnerability detection. \looseness=-1

\vspace{1.5mm}
\misbox{Unreliable}{LLMs' vulnerability detection capabilities are on par with or worse than random guessing, indicating LLMs cannot effectively detect vulnerabilities.}
Recent studies question the practical utility of LLMs in vulnerability detection. SecLLMHolmes~\cite{ullah2024llms} evaluated GPT-4 on 15 real-world file-level vulnerabilities, achieving only a 13\% (1,0) proportion (correct on both vulnerable and patched code, refer to \S\ref{subsec:setup}). Ding et al.~\cite{ding2024vulnerability} applied CoT for function-level detection, reporting (1,0) proportions of 6.21\% (GPT-3.5) and 12.94\% (GPT-4), both below the 22.7\% random-guessing baseline. Works~\cite{yin2024multitask, steenhoek2024err, zibaeirad2024vulnllmeval} also achieved results that are close to or even weaker than random.  

\vspace{2mm}
\misbox{Insensitive}{In pair-wise evaluations, LLMs exhibit precision levels close to random guessing and an exceptionally high (1, 1) proportion (false positives on patched code), suggesting that they struggle to effectively differentiate patched code from its original vulnerable counterpart.}
Results from previous works on paired evaluation (evaluated on both vulnerable and patched code) suggest that LLMs face significant challenges in differentiating patched code from vulnerable code. For instance, Ding et al.~\cite{ding2024vulnerability} found that GPT-4 with CoT reasoning achieved a (1, 1) proportion of 54.26\% on paired data, translating to a precision of approximately 0.52—close to the 0.5 threshold of random guessing. This implies that in more than half of the cases, GPT-4 incorrectly classified patched functions as vulnerable. Additionally, VulnLLMEval~\cite{zibaeirad2024vulnllmeval} demonstrated that across 10 models ranging from 7B to 70B parameters, the precision for vulnerability detection at various levels (e.g., single function, multi-functions, or multi-files) only reached between 39\% and 57\%.

\vspace{1.5mm}
\misbox{Plateaued}{The vulnerability detection capabilities of LLMs exhibit only slight variations across different architectures and sizes. Moreover, as models grow in capability or scale, their ability to detect vulnerabilities does not demonstrate significant improvement.}

Steenhoek et al.~\cite{steenhoek2024err} found that the balanced accuracy (equivalent to accuracy in a balanced dataset) of all models, ranging from 7B LLMs to GPT-4-turbo, lies between 0.5 and 0.55. The performance differences among these models are minimal, and no clear scaling law with respect to model size was observed. Similarly, Yin et al.~\cite{yin2024multitask} reported that the few-shot vulnerability detection F1-scores for models ranging from 6.7B to 34B parameters fall within 0.084–0.128, with differences as small as approximately 0.04. Additionally, Khare et al.~\cite{khare2023understanding} evaluated vulnerability detection on a real-world C/C++ dataset and found that F1-scores for models ranging from 1.5B to GPT-4 hover around 0.6 (with the 1.5B model even outperforming GPT-4), leading them to conclude that ``\textit{performance does not improve with scale.}'' \looseness=-1

These three consensus points (unreliable detection, insensitivity to patches, and plateaued performance) raise important questions about the effectiveness of current evaluation practices for LLM-based vulnerability detection. They hint that existing benchmarks may not fully capture a model's reasoning ability or its understanding of underlying vulnerabilities. These observations suggest that a more comprehensive benchmark is needed: \textit{one that accounts for contextual reasoning, traces the logic behind predictions, and helps disentangle model limitations from measurement artifacts}. Such a benchmark would offer a clearer lens through which to evaluate and improve LLM-based vulnerability detection.

\section{Design of \framework}
\label{sec:context_dataset}

We propose a novel evaluation framework for LLM-based vulnerability detection, named \framework (\textit{Context-Rich Reasoning Evaluation of Code with Trust}), designed to mitigate evaluation flaws caused by UO(I) and UO(II) (whether conclusive or inferential) that may arise due to limited context.
The core idea is to comprehensively gather as much relevant contextual information as possible to support the LLM during the detection process. Subsequently, we leverage vulnerability information to focus specifically on evaluating ground-truth vulnerabilities within the appropriate context.
Our framework operates through three stages: (1) \textbf{Vulnerability Code Context Establishment} collects extensive code dependencies and vulnerability-specific information; (2) \textbf{Context-Rich Prompts Construction} employs dual prompts (vulnerability assessment and rationale evaluation) incorporating code, patches, and contextual data; and (3) \textbf{Assessment Generation} applies standardized metrics with two evaluation modes—Lenient Mode (accepting approximate detections) and Strict Mode (enforcing precise vulnerability attribution through feedback loops). Note that \framework \textbf{serves as an evaluation framework rather than a detection tool}, systematically analyzing LLMs' reasoning capabilities in security contexts. \looseness=-1

\begin{figure*}[t] 
    \centering
    \includegraphics[width=\textwidth]{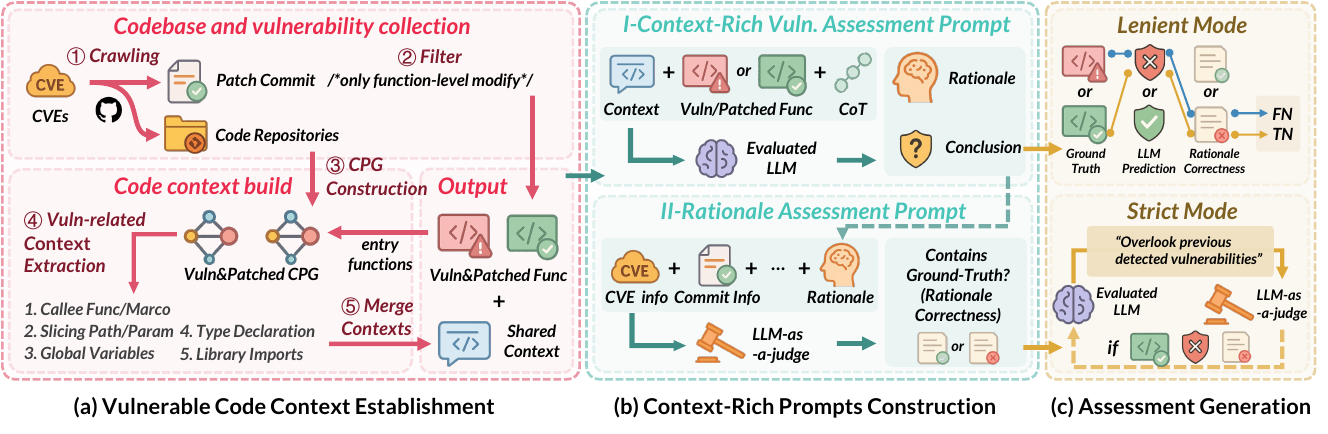} 
    \caption{The three stages of the \framework are as follows: The constructed functions and shared context are used in (b)-I; the rationales generated in (b)-I are evaluated for correctness by (b)-II, which employs LLM-as-a-judge. Finally, the assessment is conducted in (c). In (c)-Lenient Mode, only the key outcomes are displayed in this figure, with further details provided in \S\ref{subsec:assessment}.}
    \label{fig:framework}
\end{figure*}

\subsection{Vulnerable Code Context Establishment}
\label{subsec:codeestablishment}

Vulnerable code context establishment consists of two phases: (i) codebase and vulnerability collection and (ii) code context build. First, we gather vulnerable code (as well as the corresponding vulnerability information) and patched function-level code to enable side-by-side comparison. Then, we construct rich contexts for each function by extracting relevant preconditions and execution logic using slicing and code property graphs. 

\vspace{1mm}
\noindent\textbf{(Phase I) Codebase and Vulnerability Collection.} 
We collect not only the code base containing the vulnerability (as well as the vulnerability information, such as CVE descriptions) but also the corresponding patched version following the well-established practice (e.g., ~\cite{ullah2024llms, ding2024vulnerability}). The \textit{key insight} behind it stems from existing research showing that LLMs often struggle to distinguish between vulnerable code and its patched (non-vulnerable) counterpart~\cite{risse2024uncovering, ding2023concord}, as stated in Consensus \#2.
Thus, paired evaluation helps investigate this issue.

Specifically, given a CVE record and patch commit URL, we first clone the repository and extract the patch diff (\ding{192}); Then, we filter commits to retain only function-level modifications (\ding{193}), consistent with prior work~\cite{steenhoek2024err, ding2024vulnerability, yin2024multitask}; Finally, we extract vulnerable/patched function pairs from both repository versions as LLM inputs.
Structured vulnerability information (CVE IDs, CWE categories, and descriptions) is extracted from authoritative sources like NVD~\cite{NIST-NVD-2019}.
Additionally, the commit messages are collected to describe vulnerabilities and patches. \looseness=-1

\vspace{1mm}
\noindent\textbf{(Phase II) Code Context Build.} 
We prepare the context for both vulnerable and patched code. The context of a function is categorized into three types: precondition (conditions that must hold before execution), execution logic (statements involved during execution), and postcondition (conditions that must hold after execution). To be more specific: 

\begin{itemize}[partopsep=2pt, topsep=-\parskip, parsep=2pt, itemsep=2pt, leftmargin=*]
  
\item \textbf{Precondition:} 
Preconditions define the conditions required for correct function execution (e.g., whether a variable can be \texttt{NULL}), which is crucial for accurate vulnerability detection. 
Without them, LLMs might assume a ``may analyze'' scenario where all external inputs could be \texttt{NULL}, leading to numerous false positives in detecting null pointer dereferences.
While precondition inference remains an active research area, existing methods~\cite{padhi2016data, astorga2018preinfer, menguy2022automated} face scalability limitations due to their reliance on runtime traces and struggle with complex vulnerability contexts. 
To address incomplete preconditions, we adopt a lightweight solution: \textit{marking unrelated variables}. Specifically, during prompt construction, we explicitly mark unrelated function parameters (those without data/control-flow influence on vulnerable statements) as irrelevant, effectively curbing the LLM's spurious speculation and reducing false positives caused by a lack of context.

\item \textbf{Execution Logic:} 
Execution logic comprises all statements executed during function execution, including callee functions and macros for transparency, along with type declarations and global variable types, as summarized in Work~\cite{risse2024top} and illustrated in \autoref{fig:code}(b). However, large projects with deep call hierarchies often produce excessive context sizes that exceed LLM limitations.  
To balance comprehensiveness and feasibility, we implement three optimizations: 
(1) \textbf{backward/forward slicing} to retain only vulnerability-relevant statements, (2) \textbf{exclusion of standard library definitions} (e.g., \texttt{printf}) presumed to be within the LLM's knowledge base, and (3) \textbf{a two-layer callee depth restriction}—empirically validated to suffice for most cases—that captures immediate callees and their direct descendants.

\item \textbf{Postcondition:} 
Postconditions define expected states after function execution (e.g., ensuring non-\texttt{NULL} return values to prevent null pointer dereferences in calling functions). Like preconditions, they cannot be fully captured through static analysis. We exclude postconditions from our approach because: (1) their vulnerability status often depends on developer expectations rather than code semantics; (2) related vulnerabilities typically manifest externally; and (3) manual analysis of 50 patches revealed only 2 cases (4\%) requiring postcondition consideration.

\end{itemize}

Following our established guidelines, we extract vulnerability-related program elements and annotate relevant parameters using Code Property Graphs (CPGs) - unified graph representations capturing program syntax, control flow, and data dependencies (\ding{194}). To optimize resource usage for large projects (e.g., Linux kernel), we: (1) select vulnerable/patched functions as entry points for initial file filtering, creating a focused subproject; (2) construct CPGs for this subset; and (3) perform graph slicing to extract execution contexts along slicing paths (\ding{195}), including: callee functions/macros, global variables, type declarations, library imports, and slicing path parameters (statement indices and related parameters).
We extract and merge contexts from both vulnerable and patched versions (\ding{196}), then remove the modified functions (which are exclusively internal functions per our commit selection criteria) to isolate the unchanged portion common to both versions - the \textbf{Shared Context}.
This yields three key components: vulnerable functions, patched functions, and their shared context. For evaluation, we construct paired prompts using the vulnerable/patched functions to assess the LLM's detection capability.

\subsection{Context-Rich Prompts Construction}
\label{subsec:promptconstruction}
This stage generates two core evaluation prompts: (1) The \textit{context-rich vulnerability assessment prompt} combines target functions with shared context (\S\ref{subsec:codeestablishment}) to test the LLM's vulnerability identification accuracy against ground-truth, while simultaneously establishing the contextual basis for subsequent analysis; and (2) The \textit{rationale assessment prompt}, constructed post-detection, incorporates vulnerability information and the LLM's initial rationale to evaluate whether the explanation correctly identifies (or falsely alarms) the ground-truth vulnerability.

\vspace{1mm}
\noindent\textbf{(Prompt I) Context-Rich Vulnerability Assessment Prompt.}
This prompt evaluates LLM vulnerability detection through rationale generation. It includes: (1) \textbf{CWE descriptions} for specialized vulnerability detection, following~\cite{ullah2024llms, steenhoek2024err}; (2) \textbf{Context-Rich Code} prepared in \S\ref{subsec:codeestablishment}, including callee functions/macros, global variables, type declarations, library imports, and slicing paths/parameters; (3) \textbf{Assumptions} restricting analysis to marked the slicing parameters and statements with the comment ``\texttt{//potential}''; and (4) \textbf{Instructions} for zero-shot CoT reasoning to output \texttt{HAS\_VUL}/\texttt{NO\_VUL} conclusions.
We adopt zero-shot CoT because: (a) it avoids example selection biases present in few-shot approaches~\cite{steenhoek2024err}, and (b) advanced prompting techniques prove ineffective for SOTA LLMs~\cite{wang2024advanced}. For each vulnerability, we generate paired prompts (vulnerable/patched versions) to obtain detection rationales and conclusions. See Appendix \ref{appendix:prompt} for the detailed prompt.

\vspace{1mm}
\noindent\textbf{(Prompt II) Rationale Assessment Prompt.}
This prompt evaluates the correctness of LLM-generated rationales from Prompt I's preliminary results, with the evaluation being conducted by another model acting as an LLM-as-a-judge. Our prompt incorporates the initial rationale alongside ground-truth evidence (CVE description, CWE classification, commit message, and code diff) to assess alignment with known vulnerabilities (see Appendix~\ref{appendix:prompt} for structure). Crucially, our framework implements contrastive capability assessment by evaluating both vulnerable and patched versions: for vulnerable functions, the rationale must correctly identify the root cause. If multiple causes are provided by the LLM, it is acceptable as long as the root cause is included, since other causes may result from UO(II) and cannot be judged as unreasonable. For patched versions, the rationale must avoid false alarms — the LLM should not report a vulnerability for the ground-truth patched vulnerability. Reporting additional causes is allowed, as they may arise from UO(I) and also cannot be judged as unreasonable. This dual evaluation ensures that the LLM understands vulnerability distinctions, addressing our core research challenge. For both cases, the structure of the prompt remains consistent except for the instruction. The success criteria also differ appropriately: correct vulnerability attribution for vulnerable code and correct non-flagging for patched code. \looseness=-1

\subsection{Assessment Generation}
\label{subsec:assessment}
\framework produces final assessments using standardized metrics with two evaluation modes: (1) \textbf{Lenient Mode} accepts vulnerability detections with minor rationale inaccuracies, while (2) \textbf{Strict Mode} employs feedback mechanisms to ensure correct vulnerability attribution, particularly for patched functions. Formally, we represent outcomes as $(r_v, r_p, r_r) \in \{0,1\}^3$, where $r_v$ indicates ground-truth status (1=vulnerable, 0=patched), $r_p$ captures the LLM's classification ($1$=vulnerable, $0$=non-vulnerable), and $r_r$ denotes rationale correctness (T/F). Evaluation output is shown in \autoref{tab:outcome}. \looseness=-1

\begin{table}[t]
\centering
\caption{Evaluation outcomes under lenient and strict modes.}
\vspace{-0.2cm}
\footnotesize % 使用更小的字体
\setlength{\tabcolsep}{3pt} % 减少列间距
\begin{tabular}{ccccc}
\toprule
\makecell{\textbf{Ground}-\textbf{Truth}} & \makecell{\textbf{LLM} \\ \textbf{Prediction}} & \makecell{\textbf{Rationale}\\ \textbf{Correctness}} & \makecell{\textbf{Lenient} \textbf{Mode}} & \makecell{\textbf{Strict} \textbf {Mode}} \\
\textbf{($r_v$)} & \textbf{($r_p$)} & \textbf{($r_r$)} & \textbf{Output} & \textbf{Output} \\
\midrule
1 (Vul.)  & 1 (Vul.)  & T     & TP   & TP \\
1 (Vul.)  & 1 (Vul.)  & F     & \textbf{FN}   & \textbf{FN} \\
1 (Vul.)  & 0 (Pat.) & –     & FN   & FN \\
0 (Pat.) & 0 (Pat.) & –     & TN   & TN \\
0 (Pat.) & 1 (Vul.)  & T     & FP   & FP \\
0 (Pat.) & 1 (Vul.)  & F     & \textbf{TN}   & \textbf{\textit{Feedback Loop}} \\
\bottomrule
\end{tabular}
\label{tab:outcome}
\vspace{-0.5cm}
\end{table}

\begin{itemize}[partopsep=2pt, topsep=-\parskip, parsep=2pt, itemsep=2pt, leftmargin=*]
\item \textbf{Lenient Mode} applies a revision function $\delta(\cdot)$ to raw LLM outputs: (1) For vulnerable code, $\delta(\text{Vulnerable}, 1, T) = 1$ represents True Positive (i.e., correct detection and rationale), otherwise False Negatives; (2) For patched code, $\delta(\text{Patched}, 0, -) = 0$ denotes True Negative, which means current non-detections, whereas $\delta(\text{Patched}, 1, T) = 1$ marks False Positives. Notably, $\delta(\text{Patched}, 1, F) = 0$ also yields a True Negative since the rationale doesn't reference the original vulnerability.
\item \textbf{Strict Mode} enhances Lenient Mode by addressing a critical edge case: when models falsely flag patched functions as vulnerable with unrelated rationales (i.e., $\delta(\text{Patched}, 1, F)$). While Lenient Mode counts these as True Negatives, Strict Mode implements a feedback loop providing corrective guidance (instructing the model to ignore previously inferred non-ground-truth vulnerabilities and re-analyze) until the model either (1) incorrectly identifies the patched ground-truth vulnerability as vulnerable, or (2) reaches $\textsf{max\_feedback\_rounds}$ with no false alarm. Specifically, this iterative refinement prevents the LLM from being overestimated when it identifies multiple potential vulnerabilities in the code, by avoiding cases where a prominent non-ground-truth vulnerability dominates its judgment on the ground-truth vulnerability.\looseness=-1
\end{itemize}

\section{Evaluation}

As discussed in \S\ref{subsec:rq}, we presented three consensuses regarding LLM-based vulnerability detection. We believe these consensus points may not fully reflect the true capabilities of LLMs in detecting vulnerabilities due to the lack of context. Therefore,  we have formulated three research questions (RQ). We now explore these questions in detail: 
\vspace{2mm}
\begin{itemize}
\item [\textbf{RQ1.}] Is the performance of LLM-based vulnerability detection really as poor as random guessing? 
\item [\textbf{RQ2.}] Are LLMs genuinely unable to distinguish between patched code and vulnerable code? 
\item [\textbf{RQ3.}] Can scaling up the LLMs truly not improve their ability to detect vulnerabilities? 
\end{itemize}

\subsection{Experiment Setup}
\label{subsec:setup}
\vspace{2mm}
\noindent\textbf{Dataset.}
We create a dataset, which includes 2,000 program pairs (each with a vulnerable and patched version) drawn from 364 real-world projects and spanning 99 CWEs. On average, the shared context length is 13,249 tokens, and function-level code averages 1,574 tokens. See Appendix~\ref{appendix:dataset} for details. It builds on three high-quality sources: \textsf{MoreFixes}~\cite{akhoundali2024morefixes}, \textsf{PrimeVul}~\cite{ding2024vulnerability}, and \textsf{ReposVul}~\cite{wang2024reposvul}. We retrieved source projects, constructed call graphs with cflow~\cite{gnu_cflow_2021}, and used Joern~\cite{joern_bug_hunters_workbench_2024} to extract Code Property Graphs. Manual auditing of 50 pairs confirmed 98\% label accuracy. For targeted analysis, we created a CWE-1000 subset (Appendix~\ref{appendix:cwe1000}), sampling 50 pairs per top-level CWE. Due to limited data, three categories (CWE-693, CWE-697, and CWE-435) contain 30, 10, and 10 pairs, respectively, for a total of 400 pairs across 10 top-level CWEs.

\begin{figure*}[t] % 

    \centering
    \includegraphics[width=\textwidth]{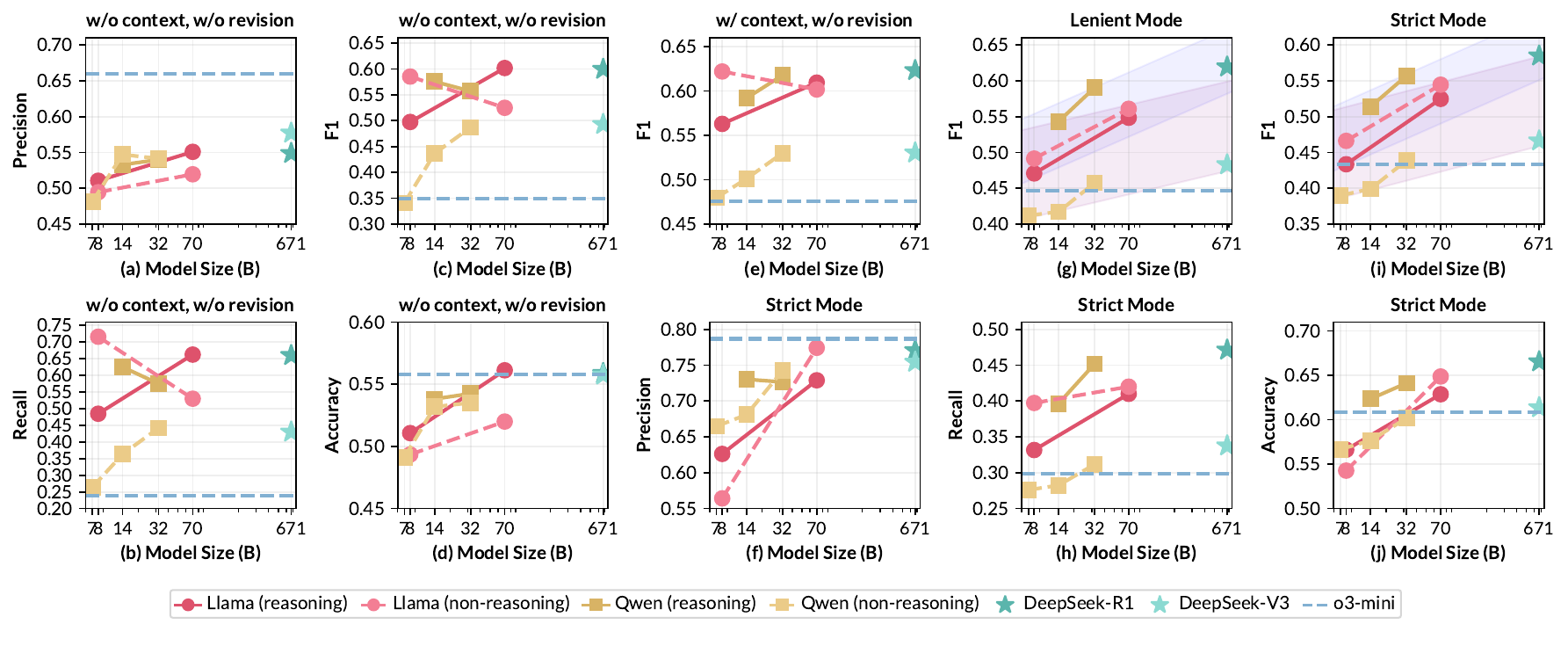} % 图片宽度设为文本宽度
    \vspace{-0.65cm}
    \caption{Performance comparison across various evaluation settings. (a)-(d): Different metrics in the ``w/o context, w/o revision'' configurations. (f), (h), (i), (j): Different metrics under Strict Mode. The model r1-qn-7b has been excluded from the analysis due to its excessive abnormal outputs, which substantially impair the clarity and reliability of the results.}
    \label{fig:f1}
\end{figure*}

\vspace{2mm}
\noindent\textbf{Model Selection.}
We next outline our strategy for selecting and configuring LLMs in the experiments. Table~\ref{tab:model_selection} presents our curated set of 13 LLMs spanning 4 model families. These models were selected based on three criteria: (1) Availability of both reasoning and non-reasoning variants, (2) Coverage of diverse parameter scales, and (3) Representation of architectural diversity. In addition to the reasoning and non-reasoning models from the Qwen, Llama, and DeepSeek series, we also evaluated the o3-mini model because it offers three reasoning modes (low, medium, and high) to control the length of the rationale. For RQ1–RQ2, we used the default o3-mini-medium configuration. For RQ3, we leveraged all three modes to investigate test-time scaling effects on vulnerability detection performance. We select GPT-4o to produce the rationale assessment result, and GPT-4o is n SOTA model widely used in LLM-as-a-judge tasks~\cite{tan2024judgebench}.
To ensure experimental stability and reproducibility, we set the temperature of all models to 0, following~\cite{ullah2024llms}. \looseness=-1

\begin{table}[t]
    \centering
    \scriptsize 
    \renewcommand{\arraystretch}{1.15}
    \caption{Models Evaluated. Abbr. denotes the abbreviated names of the models used in this paper. DS-R1-Dist refers to DeepSeek-R1-Distill, and Inst stands for Instruction.}
    \vspace{-0.3cm}
    \resizebox{\linewidth}{!}{%
    \footnotesize
    \begin{tabular}{ll ll r}
        \toprule
        \multicolumn{2}{c}{\textbf{Non-reasoning LLM}} & \multicolumn{2}{c}{\textbf{Reasoning LLM}} & \textbf{Param} \\
        Model & Abbr. & Model & Abbr. &  \\
        \midrule
        \multicolumn{5}{l}{\textbf{Qwen Series}} \\
        Qwen2.5-7B-Inst   & \textbf{qn-7b}   & DS-R1-Dist-Qwen-7B   & \textbf{r1-qn-7b}   & 7B  \\
        Qwen2.5-14B-Inst  & \textbf{qn-14b}  & DS-R1-Dist-Qwen-14B  & \textbf{r1-qn-14b}  & 14B \\
        Qwen2.5-32B-Inst  & \textbf{qn-32b}  & DS-R1-Dist-Qwen-32B  & \textbf{r1-qn-32b}  & 32B \\
        \midrule
        \multicolumn{5}{l}{\textbf{Llama Series}} \\
        Llama-3.1-8B-Inst  & \textbf{lm-8b}   & DS-R1-Dist-Llama-8B   & \textbf{r1-lm-8b}   & 8B  \\
        Llama-3.3-70B-Inst & \textbf{lm-70b}  & DS-R1-Dist-Llama-70B  & \textbf{r1-lm-70b}  & 70B \\
        \midrule
        \multicolumn{5}{l}{\textbf{DeepSeek Series}} \\
        DeepSeek-V3       & \textbf{ds-v3}   & DeepSeek-R1             & \textbf{ds-r1}     & 671B\\
        \midrule
        \multicolumn{5}{l}{\textbf{OpenAI Series}} \\
        -           & \textbf{-} & o3-mini                 & \textbf{o3-mini}   & -   \\
        \bottomrule
    \end{tabular}
    }
    \label{tab:model_selection}
    \vspace{-0.5cm}
\end{table}

\vspace{2mm}
\noindent\textbf{Metrics.}
We employ two categories of metrics to evaluate LLM capabilities in vulnerability detection. \textbf{The first category}, Conventional Metrics, includes F1-score, accuracy, recall, and precision, calculated using True Positives, False Positives, True Negatives, and False Negatives. \textbf{The second category}, Pair-wise Prediction Proportion, assesses the LLMs' ability to distinguish vulnerable functions from patched ones using pair-wise prediction evaluation, as introduced by Ding et al. (2024)~\cite{ding2024vulnerability}. Formally, let $(p_v, p_f) \in \{0,1\}^2$ denote prediction outcomes, where $p_v$ or $p_f$ indicate whether the vulnerable or patched function is detected as vulnerable (1) of not (0). Four possible outcomes exist: $(1, 0)$ for ideal detection (vulnerability identified in the original but not patched version), $(1, 1)$ for a false alarm on patched code, $(0, 0)$ for a missed vulnerability in the original function, and $(0, 1)$ for a dual misprediction. The proportion of $(1, 0)$ pairs quantifies the model's discriminative capability between vulnerable and patched code variants. In Strict Mode, \textsf{max\_feedback\_rounds} is set to 4, preventing the LLM from disregarding ground-truth vulnerabilities (see Appendix \ref{appendix:feedback}). \looseness=-1

\subsection{Detection Performance (RQ1)}

\begin{figure*}[t] %
    \centering
    \includegraphics[width=\textwidth]{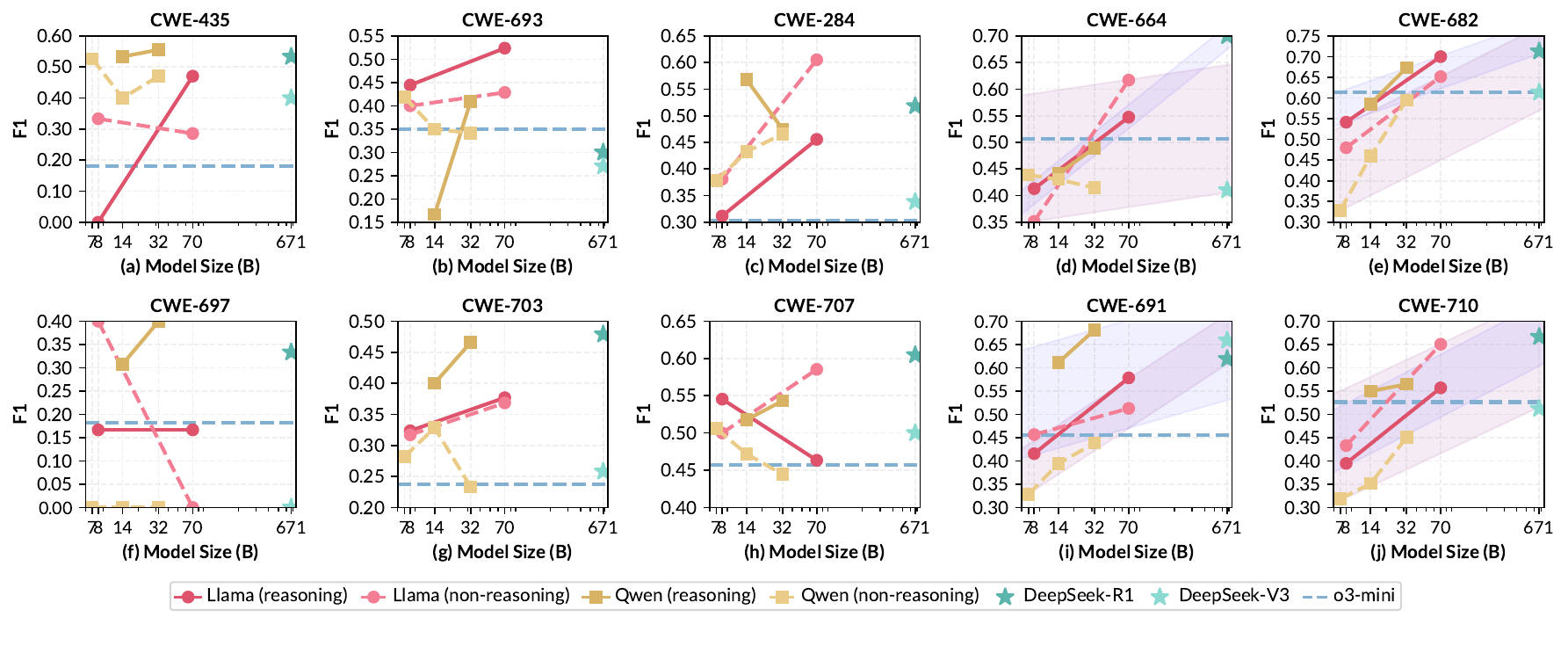} %
    \vspace{-0.7cm}
    \caption{F1-score comparison of 10 top-level CWEs.}
    \label{fig:cwe}
\end{figure*}

\vspace{2mm}
\noindent\textbf{Methodology and Settings:}
We evaluate the vulnerability detection performance of all 13 LLMs on the selected 400 pairs (\S\ref{subsec:setup}) and separately present the results for: (1) w/o context, w/o revision, which refers to function-level vulnerability detection only, with evaluation conducted on binary results, consistent with most prior works~\cite{ding2024vulnerability, steenhoek2024err, yin2024multitask}. (2) w/ context, w/o revision, which involves utilizing datasets containing contextual information (\S\ref{subsec:codeestablishment}) without undergoing the revision process (\S\ref{subsec:assessment}). (3) Lenient Mode. (4) Strict Mode. The final results, including F1-score, accuracy, and other metrics, are displayed in \autoref{fig:f1}, while the pair-wise prediction proportions are shown in \autoref{fig:pair-wise}. Results for r1-qn-7b are omitted in \autoref{fig:f1} due to anomalies causing poor performance, explained in Appendix \ref{appendix:abnormaloutput}. 
Vulnerability types are presented in \autoref{fig:cwe} and \autoref{tab:cwe-data}. \autoref{fig:cwe} illustrates the F1-scores of the 10 top-CWE across various models, while \autoref{tab:cwe-data} provides an overview of these CWEs, including their brief descriptions (with full details available in Appendix~\ref{appendix:cwe1000}) and the maximum F1-score achieved. \autoref{tab:cwe-data} also presents the prevalence of each CWE type in real-world datasets: MoreFixes~\cite{akhoundali2024morefixes}, PrimeVul~\cite{ding2024vulnerability}, and ReposVul~\cite{wang2024reposvul}. These proportions reflect their occurrence in real-world scenarios.

\vspace{2mm}
\noindent\textbf{Results:}
To demonstrate that the \misref{Unreliable} in previous works arose from incorrect evaluations due to the lack of context, we first analyze the results obtained under the w/o context, w/o revision setting (consistent with prior works). As shown in \autoref{fig:f1}(c) and (d), most models achieved F1-scores ranging from 0.5 to 0.6 and 0.5 to 0.55 accuracy, which is consistent with earlier studies~\cite{steenhoek2024err, khare2023understanding} that state LLM-based vulnerability detection performs close to random guessing. 

When using \framework to minimize the impact of missing context, the results from \autoref{fig:f1}(i) and (j) reveal that SOTA models like DeepSeek-R1 achieved an F1-score of 0.6 and 67\% accuracy in Strict Mode, with smaller models reaching approximately 60\% accuracy. Additionally, \autoref{fig:pair-wise} shows that DeepSeek-R1 achieved a maximum (1, 0) proportion of 37\%, significantly surpassing the 25\% random baseline. 
Notably, the vulnerability detection task in \framework is more rigorous compared to previous binary-only tasks. This is because \framework not only requires the predicted label to be correct but also demands precise reasoning about the underlying causes of the vulnerabilities. Despite this increased level of difficulty and comprehensiveness, the models demonstrated higher vulnerability detection capabilities than previously reported, surpassing the random baseline. This suggests:

\vspace{2mm}
\findbox{The vulnerability detection capabilities of LLMs have been underestimated due to invalid evaluations stemming from a lack of context. \framework demonstrates that SOTA models, such as DeepSeek-R1, achieve up to 67\% accuracy and a 37\% (1, 0) proportion on more rigorous tasks—\\significantly surpassing the random baseline.}
\vspace{2mm}

Steenhoek et al.~\cite{steenhoek2024err} further analyzed the performance of LLM-based vulnerability detection across different vulnerability types. The results revealed that detection outcomes for all types were nearly random, raising the question of whether mitigating context limitations could enable LLMs to achieve better performance across all vulnerability types.

\begin{table}[h]
    \centering
    \caption{Evaluation results across different CWEs.}
    \label{tab:cwe-data}
    \vspace{-0.3cm}
    \setlength{\tabcolsep}{2.5pt}
    \renewcommand{\arraystretch}{1.15}
    \footnotesize
    \renewcommand{\arrayrulewidth}{0.3pt} 
    \begin{tabular}{
        l %
        c %
        l %
        r %
        r % 
    }
        \toprule
        \textbf{Type} & \textbf{CWE} & \textbf{Description} & {\textbf{Prop.}} & {\textbf{Max F1}} \\
        \midrule
        \multirow{6}{*}[-0.8ex]{\textbf{Common}} 
            & \textbf{664} & Improper Resources Control & \textbf{56.2\%} & \textbf{\textcolor{llama-r}{0.700}} \\ 
            % \cmidrule(l){2-5}
            & \textbf{682} & Incorrect Calculation & \textbf{5.8\%} & \textbf{\textcolor{llama-r}{0.713}} \\ 
            % \cmidrule(l){2-5}
            & \textbf{691} & Insufficient Control Flow Management & \textbf{5.8\%} & \textbf{\textcolor{llama-r}{0.681}} \\ 
            % \cmidrule(l){2-5}
            & \textbf{710} & Improper Adherence to Coding Standards & \textbf{7.7\%} & \textbf{\textcolor{llama-r}{0.667}} \\ 
            \cmidrule(l){2-5}
            & \textbf{284} & Improper Access Control & \textbf{6.8\%} & \textbf{{0.605}} \\ 
            % \cmidrule(l){2-5}
            & \textbf{707} & Improper Neutralization & \textbf{7.3\%} & \textbf{{0.605}} \\ 
        \midrule
        \multirow{4}{*}[-0.8ex]{\textbf{Rare}} 
            & \textbf{435} & Improper Multi-Entity Interaction & \textbf{\textcolor{qwen-r}{0.5\%}} & \textbf{\textcolor{deepseek-r}{0.556}} \\ 
            & \textbf{693} & Protection Mechanism Failure & \textbf{\textcolor{qwen-r}{2.5\%}} & \textbf{\textcolor{deepseek-r}{0.524}} \\ 
            & \textbf{697} & Incorrect Comparison & \textbf{\textcolor{qwen-r}{0.5\%}} & \textbf{\textcolor{deepseek-r}{0.400}} \\ 
            \cmidrule(l){2-5}
            & \textbf{703} & Improper Check of Exceptional Conditions & \textbf{7.0\%} & \textbf{\textcolor{deepseek-r}{0.479}} \\ 
        \bottomrule
    \end{tabular}
\end{table}

To further classify vulnerabilities, we divided the ten top-level CWE categories into two groups based on prevalence: \textit{Common} and \textit{Rare}, as shown in \autoref{tab:cwe-data}. Vulnerabilities classified as \textit{Rare} either occur infrequently in real-world scenarios (CWE-435, 693, 697) or involve rare patterns (CWE-703, which involves handling rare exceptional conditions). The remaining six categories, representing frequently observed vulnerabilities, were labeled as \textit{Common}.

The findings in \autoref{tab:cwe-data} indicate: (1) LLMs perform significantly worse on \textit{Rare} vulnerabilities compared to \textit{Common} ones. (2) Within the \textit{Common} group, performance varies. Vulnerabilities with simpler, fixed patterns—such as CWE-664 (e.g., out-of-bounds read), CWE-682 (e.g., integer overflow), CWE-691 (e.g., infinite loops), and CWE-710 (e.g., NULL pointer dereference)—achieve high F1-scores of around 0.7 with SOTA models. In contrast, vulnerabilities tied to complex application scenarios and system security, such as CWE-284 and CWE-707, show weaker detection performance.
\newpage

\vspace{2mm}
\findbox{When context-related challenges are resolved, LLMs show varying effectiveness across vulnerability types, outperforming random baselines. They excel at detecting common, fixed-pattern vulnerabilities, achieving F1-scores around 0.7, but struggle with rare vulnerabilities, highlighting a key limitation in generalization.}

\subsection{Code Distinguishability (RQ2)}

\begin{figure}[t] %
    \centering
    \hspace{-0.5cm}
    \includegraphics[width=0.46\textwidth]{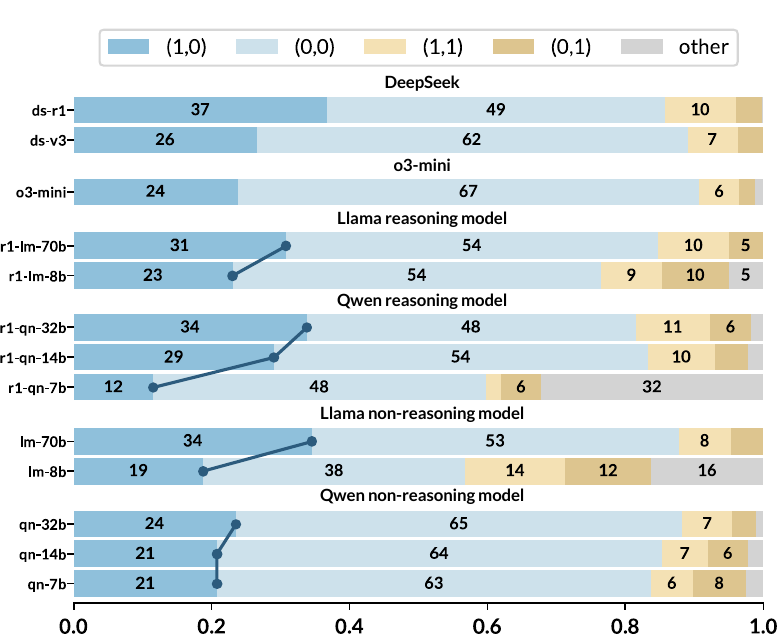} % 图片宽度设为文本宽度
    \vspace{-0.3cm}
    \caption{Pair-wise prediction proportion of all LLMs.}
    \label{fig:pair-wise}
\end{figure}

\vspace{2mm}
\noindent\textbf{Methodology and Settings:}
Initially, we will analyze the overall statistical data obtained in RQ1 to examine the (1, 1) proportion and precision of LLMs in vulnerability detection. However, since numerical analysis alone may not be sufficiently convincing, we will further investigate and categorize the reasons behind LLM reasoning errors. \looseness=-1

To investigate why models tend to incorrectly classify patched code as vulnerable, while also distinguishing the capabilities of non-reasoning and reasoning models, we selected the non-reasoning LLM ds-v3 and the reasoning LLM ds-r1. We manually inspected cases where the two models produced inconsistent results on patched code (i.e., False Positive (FP) versus True Negative (TN) predictions). The results are presented in \autoref{tab:diff}, where R represents the reasoning LLM (ds-r1), and Non-R represents the non-reasoning LLM (ds-v3).

\vspace{2mm}
\noindent\textbf{Results:}
As for \misref{Insensitive}, in the w/o context, w/o revision setting, the precision of nearly all models is concentrated between 0.5 and 0.55 (\autoref{fig:f1} (a)), which is close to the random baseline of 0.5. This observation aligns with previous works stating that LLMs cannot effectively distinguish patched code from its corresponding vulnerable code.
After evaluating using \framework, \autoref{fig:f1}(f) shows that larger models achieve notably higher precision. Specifically, the 671B-parameter SOTA models, such as ds-r1 and ds-v3, reach a precision close to 0.8, indicating that these models are unlikely to misclassify patched functions as still vulnerable. Additionally, as shown in \autoref{fig:pair-wise}, all models exhibit a (1, 1) proportion of approximately 10\%, which is significantly lower than the random baseline.

\vspace{1.5mm}
\findbox{The absence of sufficient context leads to an ineffective evaluation of the model's precision and its (1, 1) proportion, which represents its capability to differentiate between patched code and the associated original vulnerable code. SOTA models have demonstrated strong performance in this area, achieving precision rates close to 0.8.}

\begin{table}[t]
    \centering
    \caption{Attribution of differences for TN $\leftrightarrow$ FP between non-reasoning LLM (Non-R) ds-v3 and reasoning LLM (R) ds-r1.\looseness=-1}
    \vspace{-0.3cm}
    \label{tab:diff}
    % \begin{subtable}[t]{\linewidth}
    %     \centering
    %     \caption{Attribution of differences for TP $\rightarrow$ FN transitions.}
    %     \begin{tabular}{lcc}
    %         \toprule
    %         \multirow{2}{*}{\textbf{Reason}} & \multicolumn{2}{c}{\textbf{TP $\rightarrow$ FN}} \\
    %         \cmidrule(lr){2-3}
    %          & \textbf{\cmark R$\rightarrow$ \xmark NR} & \textbf{\cmark NR$\rightarrow$ \xmark R} \\
    %         \midrule
    %         Vuln. Not Found         & 36 & 2 \\
    %         Vuln. Reasoning Error   & 43 & 17 \\
    %         \quad - Minimum Reasoning & 17  & 0 \\
    %         \quad - Procedural Error & 26 & 8 \\
    %         \quad - Mis-Corrected Reasoning & 0 & 9 \\
    %         \bottomrule
    %     \end{tabular}
    % \end{subtable}
    
    % \vspace{1em} % Space between subtables
    
    % Subtable 2
    \begin{subtable}[t]{\linewidth}
        \centering
        % \caption{Attribution of differences for TN $\rightarrow$ FP transitions.}
        \footnotesize
        \begin{tabular}{lcc}
            \toprule
            \multirow{2}{*}{\textbf{Reason}} & \multicolumn{2}{c}{\textbf{\textcolor{deepseek-r}{TN} $\leftrightarrow$ \textcolor{llama-r}{FP}}} \\
            \cmidrule(lr){2-3}
             & \textbf{\textcolor{deepseek-r}{R\ }$\leftrightarrow$\textcolor{llama-r}{\ Non-R}} & \textbf{\textcolor{deepseek-r}{Non-R\ }$\leftrightarrow$\textcolor{llama-r}{\ R}} \\
            \midrule
            Patch Ignored           & 6  & 2 \\
            Patch Deemed Insufficient & 10 & 19 \\
            \quad - Minimum Reasoning & 2  & 0 \\
            \quad - Procedural Error & 8  & 13 \\
            \quad - Mis-Corrected Reasoning & 0 & 6 \\
            \bottomrule
        \end{tabular}
    \vspace{-0.4cm}
    \end{subtable}
\end{table}

Next, we will conduct a detailed analysis of cases to attribute errors. Errors in \autoref{tab:diff} are classified into two main types. The first type, \textit{Patch Ignored}, occurs when the rationale does not mention the patch at all, suggesting that the LLM failed to notice its presence. Only this scenario can effectively highlight the flaw in LLM's inability to distinguish, where for ``similar but patched'' code, the inability to distinguish means that the LLM is unaware of the existence of the patch, i.e., it considers the patched code to be the same as before the patch.
The second type, \textit{Patch Deemed Insufficient}, arises when the LLM acknowledges the patch but erroneously concludes that it does not adequately resolve the vulnerability. This type can be further broken down into the following subcategories:

\begin{itemize}[partopsep=2pt, topsep=-\parskip, parsep=2pt, itemsep=2pt, leftmargin=*]
    \item \textbf{Minimum Reasoning.} The model provides no substantive analysis of the vulnerable region and instead directly asserts whether the code is vulnerable or not after merely restating it.
    \item \textbf{Procedural Error.} The model engages in reasoning but makes a mistake during the process, resulting in an incorrect conclusion.
    \item \textbf{Mis-Corrected Reasoning.} The model successfully activates reasoning, and correctly identifies the ground-truth vulnerability or patch during the process, but subsequently contradicts itself and alters the answer incorrectly.
\end{itemize}

Based on the results, it can be observed that ``Patch ignored'' accounts for only a small portion of LLM's vulnerability attribution, and the reasoning model effectively reduces its occurrence (37.5\% vs. 9.5\%). This indicates that under the premise of high LLM precision, only a small fraction of the remaining false positives (FP) are truly cases where the LLM fails to distinguish between vulnerable and non-vulnerable code. In most cases, the LLM is aware of the existence of the patch but makes an incorrect inference.

The specific reasons for these incorrect inferences differ between non-reasoning models and reasoning models. Non-reasoning models are more prone to the phenomenon of ``Minimum Reasoning,'' where the LLM uses CoT reasoning but only performs superficial reasoning, essentially describing the functionality of the code without detailed analysis of the patch, leading to incorrect outputs. We consider this to be the main flaw of non-reasoning models: even when using prompt-driven reasoning methods, they fail to consistently trigger System 2 thinking (refer to \S\ref{subsec:backgroundreasoning}). On the other hand, reasoning models are more susceptible to ``Mis-Corrected Reasoning,'' where the model, due to over-thinking, changes a correct answer to an incorrect one. This is also one of the flaws of reasoning models. See Appendix \ref{appendix:patterns} for cases of different categories.

\vspace{1.5mm}
\findbox{At very high precision levels, only a small fraction of the remaining false positives (FPs) reflect genuine cases where the LLM struggles to differentiate between similar yet patched code. Reasoning models significantly reduce this proportion. In fact, most FPs arise because the model identifies the patch but mistakenly deems it insufficient to address the vulnerability. The strength of reasoning models lies in their ability to reliably engage System 2 thinking, but their limitation is a tendency to over-thinking, sometimes turning correct answers into incorrect ones.\looseness=-1}

\subsection{Capability Enhancement (RQ3)}

\begin{figure*}[t] %

    \centering
    \includegraphics[width=\textwidth]{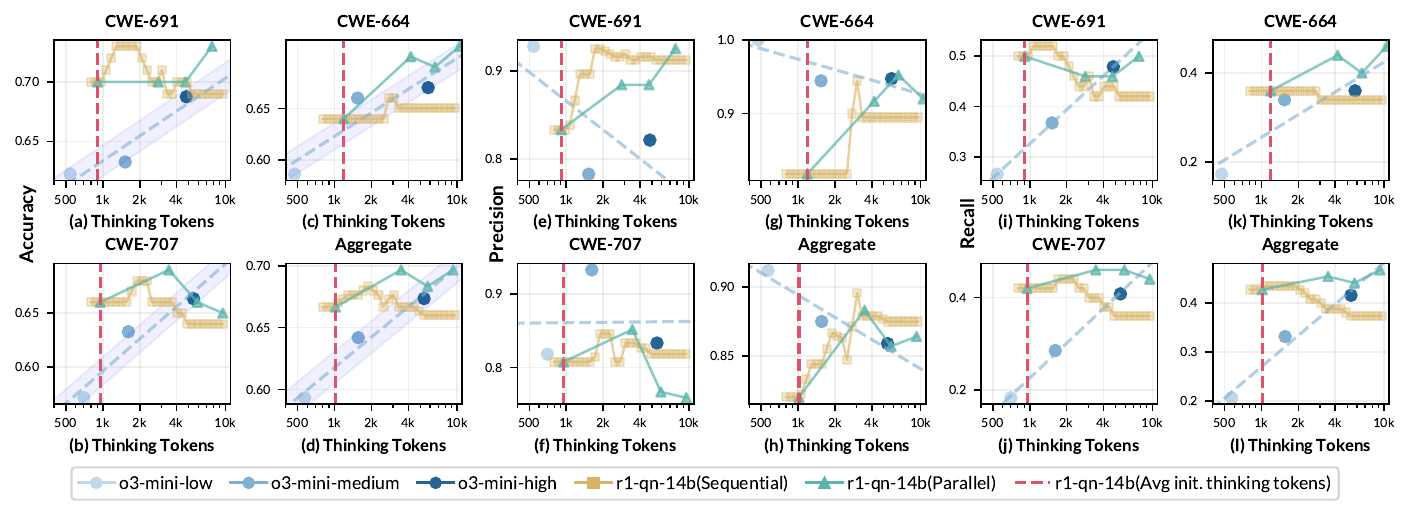} %
    \vspace{-0.7cm}
    \caption{Test-test scaling of o3-mini and r1-qn-14b. (a)–(d), (e)–(h), and (i)–(l) show accuracy, precision, and recall, respectively.}
    \vspace{-0.1cm}
    \label{fig:test-time-scaling}
\end{figure*}

\vspace{2mm}
\noindent\textbf{Methodology and Settings:} We separately study two highly discussed types of scaling laws: model size scaling~\cite{kaplan2020scaling} and test-time scaling~\cite{snell2024scaling}. For model size scaling, we analyze the overall statistical data obtained in RQ1. For test-time scaling, we separately investigate the effects of sequential scaling and parallel scaling (\S\ref{subsec:backgroundreasoning}). We selected r1-qn-14b for its ease of deployment and strong reasoning capabilities. For sequential scaling, we applied the approach from Muennighoff et al.~\cite{muennighoff2025s1}, appending ``Wait'' during the reasoning process to lengthen the output up to 10,000 tokens using greedy decoding. For parallel scaling, we adopted Universal Self-Consistency (USC), generating multiple outputs (with temperature = 0.6) and applying majority voting across 3, 5, and 8 rationales, following Chen et al.~\cite{chen2023universal}'s setup. We also tested o3-mini, which provides low, medium, and high reasoning effort options via its API, though its internal scaling method is undisclosed and thus unclassified as sequential or parallel. For dataset selection, we used three CWEs—CWE-691 (strong improvement from reasoning LLMs), CWE-664 (moderate improvement), and CWE-707 (weaker improvement)—to assess model behavior across varying difficulty levels. To simplify the statistical analysis of reasoning token length, we evaluated all models in Lenient Mode, which ensures each input yields a single rationale without feedback. The experimental results of test-time scaling are presented in \autoref{fig:test-time-scaling}, with ``Aggregate'' representing the combined outcomes across the three CWEs.

\vspace{2mm}
\noindent\textbf{Results:}
We separately present our exploration of the model size scaling law and the test-time scaling law (SL) on LLM-based vulnerability detection.

\vspace{1mm}
\noindent\textbf{(SL-I) Model Size Scaling Law:}
We first analyze whether significant differences exist in vulnerability detection performance across various models. As shown in \autoref{fig:f1}(c), under the ``w/o context, w/o revision'' scenario, most models achieve F1-scores clustered between 0.5 and 0.6, with only minor variations. Notably, the smaller r1-lm-8b model performs comparably to the SOTA ds-r1, providing no evidence of a model size scaling law—a finding consistent with prior research.
However, when evaluated using \framework, significant performance disparities emerge across models in both Lenient Mode (\autoref{fig:f1}(g)) and Strict Mode (\autoref{fig:f1}(i)). Clear model size scaling laws are observed, applicable to both reasoning and non-reasoning LLMs.
Additionally, the results in \autoref{fig:pair-wise} confirm a scaling effect: as the number of model parameters increases, the (1, 0) proportion rises notably, indicating a marked improvement in vulnerability detection performance.

From \autoref{fig:cwe}, it is evident that the four CWE types exhibiting a clear scaling law correspond precisely to the vulnerabilities with the best detection performance. These belong to the \textit{Common} category, while other vulnerability types show no significant scaling effect.

\vspace{1.5mm}
\findbox{The flawed evaluations caused by the lack of context fail to accurately reflect the true capabilities of models, resulting in minimal differences between weaker and stronger models. However, after minimizing the impact of context, LLM-based vulnerability detection reveals significant distinctions across different models, with a clear scaling law emerging as model size increases.}

\noindent\textbf{(SL-II) Test-time Scaling Law:}
Next, we delve deeper into whether the test-time scaling effect holds true for vulnerability detection in LLMs.
% \noindent\textbf{Test-test scaling of o3-mini:}
For the o3-mini model, we calculated the average thinking tokens required by o3-mini to address vulnerabilities in corresponding categories. It can be observed that the average thinking tokens for o3-mini-low, o3-mini-medium, and o3-mini-high are 500+, 1,000+, and 5,000+, respectively. Overall, the accuracy exhibits a scaling trend (\autoref{fig:test-time-scaling} (d)), indicating that test-time scaling exists in vulnerability detection on o3-mini. Specifically, accuracy exhibits an approximate power-law relationship with thinking tokens, meaning that as computational time increases exponentially, accuracy grows nearly linearly, which suggests that improving vulnerability detection through test-time scaling is not economical. For instance, o3-mini-high, with five times the thinking tokens of o3-mini-medium, achieves an accuracy improvement of less than 0.05. \looseness=-1

\vspace{1.5mm}
\findbox{A clear test-time scaling trend is observed in the o3-mini model. However, due to the power-law relationship, it is not economically viable to rely solely on increasing test time to improve accuracy.}
In \autoref{fig:test-time-scaling}, sequential scaling of r1-qn-14b exhibits an accuracy trend that first increases, reaches a peak, and then declines, eventually falling below the initial result. In contrast, parallel scaling shows that accuracy generally increases with more thinking tokens, except for CWE-707. 
Parallel scaling demonstrates superior scalability compared to sequential scaling, as it avoids performance degradation at higher thinking token counts. By analyzing precision and recall, we observe that: (i) Precision is similar for both scaling methods across 1k–4k thinking tokens. (ii) Recall, however, differs significantly: sequential scaling shows a clear downward trend due to increased model ``conservatism,'' which reduces the number of detected vulnerabilities. In contrast, parallel scaling maintains stability or even improves recall by aggregating multiple samples, thereby avoiding overthinking and incorrect modifications. 
For further discussion, refer to Appendix \ref{appendix:test-time-scaling}.
\vspace{1.5mm}
\findbox{Both sequential and parallel scaling can enhance model performance. However, sequential scaling tends to induce excessive reflection, leading to a decline in recall. In contrast, parallel scaling demonstrates superior scalability and avoids this issue.}
\vspace{-2mm}

\section{Discussion}

\vspace{1mm}
\noindent\textbf{Root Causes Analysis:} 
% The root cause of Misunderstanding \#1 and \#2 in prior work is UO(I) caused by the lack of context. 
The root cause of the misconceptions (Consensus \#1 and \#2) in prior work is UO(I) caused by the lack of context.
This primarily affects false positive (FP) judgments, with precision in the w/o context, w/o revision setting resembling random guessing, as shown in \autoref{fig:f1}(a) and \autoref{fig:f1}(f). Under Strict Mode, precision improves to around 0.8 on SOTA models. The lack of context causes ambiguity in determining if patched code is still vulnerable, leading to misleading FPs and underestimating LLMs' capabilities, resulting in Consensus \#1 and \#2. 
% Misunderstanding \#3 is caused by UO(II) and is also due to missing context.
Consensus \#3, another misconception, results from UO(II) in the absence of sufficient context.
This affects false negative (FN) judgments, with recall in the w/o context, w/o revision setting becoming unstable, as shown in \autoref{fig:f1}(b) vs. \autoref{fig:f1}(h). Smaller models may outperform SOTA models by over-predicting, leading to a bias that favors "aggressive" models. The absence of context amplifies this, making code appear more vulnerable than it is, resulting in underestimating truly capable models. In fact, recall remains below 0.5 and improves slowly with model scaling.

\vspace{1mm}
\noindent\textbf{Suggestions:}
Two strategies should be considered when designing LLM-based vulnerability detection in the future. First, \framework emphasizes the critical role of context in vulnerability detection. However, due to the complexity of real-world programs, gathering complete context is often impractical. Therefore, future research should go beyond treating programs as plain text and focus on designing vulnerability-specific context providers, which would facilitate a more integrated approach between LLMs and static or dynamic program analysis. Second, the primary challenge for LLMs lies in recall (SOTA LLM \textasciitilde 0.5), rather than precision (SOTA LLM \textasciitilde 0.8). Once LLMs can reliably identify real vulnerabilities, they are also more likely to correctly identify non-vulnerable code. Consequently, future research should prioritize improving LLMs' ability to detect vulnerabilities accurately, which demands a deeper understanding of the vulnerabilities themselves.

\vspace{1mm}
\noindent\textbf{Limitations:} Our work is not without limitations. First, \framework utilizes LLM-as-a-judge to assess whether the rationale aligns with the ground truth. While LLM-as-a-judge can understand vulnerability semantics and reduce human effort, its judgment is not infallible. To evaluate its accuracy, we randomly sampled 50 cases and reviewed all rationales requiring LLM-as-a-judge evaluation on ds-v3 and ds-r1. Using GPT-4o as the LLM-as-a-judge, we achieved an accuracy rate of 91/99, or 92\%, which we believe can truly reflect the performance of vulnerability detection. Future work could investigate advanced techniques for LLM-as-a-judge, such as majority voting, to further improve its effectiveness.
Second, our approach aims to collect as much context as possible to enhance evaluation accuracy. However, in practice, it is impossible to gather all the relevant context, especially when information may be missing for certain vulnerabilities. Despite this challenge, we are confident that our methodology provides a solid foundation for accurate vulnerability detection and analysis.
\vspace{-0.3cm}
\section{Conclusion}

This work revisits the vulnerability detection capabilities of LLM by addressing a major gap in prior evaluations: the lack of context. We introduce \framework, a context-rich evaluation framework, and show that the widely accepted beliefs (LLMs being unreliable, insensitive to patches, and plateaued in performance) stem largely from flawed, context-free benchmarks. Our evaluation across 13 LLMs and 400 program pairs reveals that, when given appropriate context, LLMs demonstrate significantly improved accuracy and reasoning ability. We call on the security and AI communities to re-evaluate current benchmarking practices. Without accounting for context, we risk underestimating both the capabilities and the limitations of LLMs.

\bibliographystyle{ACM-Reference-Format}
\bibliography{software, website}


\begin{thebibliography}{61}

%%% the Web address.

\ifx \showCODEN    \undefined \def \showCODEN     #1{\unskip}     \fi
\ifx \showDOI      \undefined \def \showDOI       #1{#1}\fi
\ifx \showISBNx    \undefined \def \showISBNx     #1{\unskip}     \fi
\ifx \showISBNxiii \undefined \def \showISBNxiii  #1{\unskip}     \fi
\ifx \showISSN     \undefined \def \showISSN      #1{\unskip}     \fi
\ifx \showLCCN     \undefined \def \showLCCN      #1{\unskip}     \fi
\ifx \shownote     \undefined \def \shownote      #1{#1}          \fi
\ifx \showarticletitle \undefined \def \showarticletitle #1{#1}   \fi
\ifx \showURL      \undefined \def \showURL       {\relax}        \fi
% invisible to TeX
\providecommand\bibfield[2]{#2}
\providecommand\bibinfo[2]{#2}
\providecommand\natexlab[1]{#1}
\providecommand\showeprint[2][]{arXiv:#2}

\bibitem[Ahn et~al\mbox{.}(2024)]%
        {ahn2024large}
\bibfield{author}{\bibinfo{person}{Janice Ahn}, \bibinfo{person}{Rishu Verma}, \bibinfo{person}{Renze Lou}, \bibinfo{person}{Di Liu}, \bibinfo{person}{Rui Zhang}, {and} \bibinfo{person}{Wenpeng Yin}.} \bibinfo{year}{2024}\natexlab{}.
\newblock \showarticletitle{Large language models for mathematical reasoning: Progresses and challenges}.
\newblock \bibinfo{journal}{\emph{arXiv preprint arXiv:2402.00157}} (\bibinfo{year}{2024}).
\newblock


\bibitem[Akhoundali et~al\mbox{.}(2024)]%
        {akhoundali2024morefixes}
\bibfield{author}{\bibinfo{person}{Jafar Akhoundali}, \bibinfo{person}{Sajad~Rahim Nouri}, \bibinfo{person}{Kristian Rietveld}, {and} \bibinfo{person}{Olga Gadyatskaya}.} \bibinfo{year}{2024}\natexlab{}.
\newblock \showarticletitle{MoreFixes: A large-scale dataset of CVE fix commits mined through enhanced repository discovery}. In \bibinfo{booktitle}{\emph{Proceedings of the 20th International Conference on Predictive Models and Data Analytics in Software Engineering}}. \bibinfo{pages}{42--51}.
\newblock


\bibitem[Astorga et~al\mbox{.}(2018)]%
        {astorga2018preinfer}
\bibfield{author}{\bibinfo{person}{Angello Astorga}, \bibinfo{person}{Siwakorn Srisakaokul}, \bibinfo{person}{Xusheng Xiao}, {and} \bibinfo{person}{Tao Xie}.} \bibinfo{year}{2018}\natexlab{}.
\newblock \showarticletitle{PreInfer: Automatic inference of preconditions via symbolic analysis}. In \bibinfo{booktitle}{\emph{2018 48th Annual IEEE/IFIP International Conference on Dependable Systems and Networks (DSN)}}. IEEE, \bibinfo{pages}{678--689}.
\newblock


\bibitem[Bhargava and Ng(2022)]%
        {bhargava2022commonsense}
\bibfield{author}{\bibinfo{person}{Prajjwal Bhargava} {and} \bibinfo{person}{Vincent Ng}.} \bibinfo{year}{2022}\natexlab{}.
\newblock \showarticletitle{Commonsense knowledge reasoning and generation with pre-trained language models: A survey}. In \bibinfo{booktitle}{\emph{Proceedings of the AAAI conference on artificial intelligence}}, Vol.~\bibinfo{volume}{36}. \bibinfo{pages}{12317--12325}.
\newblock


\bibitem[Chen et~al\mbox{.}(2023)]%
        {chen2023universal}
\bibfield{author}{\bibinfo{person}{Xinyun Chen}, \bibinfo{person}{Renat Aksitov}, \bibinfo{person}{Uri Alon}, \bibinfo{person}{Jie Ren}, \bibinfo{person}{Kefan Xiao}, \bibinfo{person}{Pengcheng Yin}, \bibinfo{person}{Sushant Prakash}, \bibinfo{person}{Charles Sutton}, \bibinfo{person}{Xuezhi Wang}, {and} \bibinfo{person}{Denny Zhou}.} \bibinfo{year}{2023}\natexlab{}.
\newblock \showarticletitle{Universal self-consistency for large language model generation}.
\newblock \bibinfo{journal}{\emph{arXiv preprint arXiv:2311.17311}} (\bibinfo{year}{2023}).
\newblock


\bibitem[Croft et~al\mbox{.}(2023)]%
        {croft2023data}
\bibfield{author}{\bibinfo{person}{Roland Croft}, \bibinfo{person}{M~Ali Babar}, {and} \bibinfo{person}{M~Mehdi Kholoosi}.} \bibinfo{year}{2023}\natexlab{}.
\newblock \showarticletitle{Data quality for software vulnerability datasets}. In \bibinfo{booktitle}{\emph{2023 IEEE/ACM 45th International Conference on Software Engineering (ICSE)}}. IEEE, \bibinfo{pages}{121--133}.
\newblock


\bibitem[Cursor(2025)]%
        {cursor_ai_editor}
\bibfield{author}{\bibinfo{person}{Cursor}.} \bibinfo{year}{2025}\natexlab{}.
\newblock \bibinfo{title}{Cursor - The AI Code Editor}.
\newblock
\newblock
\urldef\tempurl%
\url{https://www.cursor.com}
\showURL{%
\tempurl}


\bibitem[CVE.org(2025)]%
        {cve2025}
\bibfield{author}{\bibinfo{person}{CVE.org}.} \bibinfo{year}{2025}\natexlab{}.
\newblock \bibinfo{title}{CVE.org}.
\newblock
\newblock
\urldef\tempurl%
\url{https://www.cve.org/}
\showURL{%
\tempurl}


\bibitem[Ding et~al\mbox{.}(2024a)]%
        {ding2024reasoning}
\bibfield{author}{\bibinfo{person}{Hao Ding}, \bibinfo{person}{Ziwei Fan}, \bibinfo{person}{Ingo Guehring}, \bibinfo{person}{Gaurav Gupta}, \bibinfo{person}{Wooseok Ha}, \bibinfo{person}{Jun Huan}, \bibinfo{person}{Linbo Liu}, \bibinfo{person}{Behrooz Omidvar-Tehrani}, \bibinfo{person}{Shiqi Wang}, {and} \bibinfo{person}{Hao Zhou}.} \bibinfo{year}{2024}\natexlab{a}.
\newblock \showarticletitle{Reasoning and planning with large language models in code development}. In \bibinfo{booktitle}{\emph{Proceedings of the 30th ACM SIGKDD Conference on Knowledge Discovery and Data Mining}}. \bibinfo{pages}{6480--6490}.
\newblock


\bibitem[Ding et~al\mbox{.}(2023)]%
        {ding2023concord}
\bibfield{author}{\bibinfo{person}{Yangruibo Ding}, \bibinfo{person}{Saikat Chakraborty}, \bibinfo{person}{Luca Buratti}, \bibinfo{person}{Saurabh Pujar}, \bibinfo{person}{Alessandro Morari}, \bibinfo{person}{Gail Kaiser}, {and} \bibinfo{person}{Baishakhi Ray}.} \bibinfo{year}{2023}\natexlab{}.
\newblock \showarticletitle{Concord: clone-aware contrastive learning for source code}. In \bibinfo{booktitle}{\emph{Proceedings of the 32nd ACM SIGSOFT International Symposium on Software Testing and Analysis}}. \bibinfo{pages}{26--38}.
\newblock


\bibitem[Ding et~al\mbox{.}(2024b)]%
        {ding2024vulnerability}
\bibfield{author}{\bibinfo{person}{Yangruibo Ding}, \bibinfo{person}{Yanjun Fu}, \bibinfo{person}{Omniyyah Ibrahim}, \bibinfo{person}{Chawin Sitawarin}, \bibinfo{person}{Xinyun Chen}, \bibinfo{person}{Basel Alomair}, \bibinfo{person}{David Wagner}, \bibinfo{person}{Baishakhi Ray}, {and} \bibinfo{person}{Yizheng Chen}.} \bibinfo{year}{2024}\natexlab{b}.
\newblock \showarticletitle{Vulnerability Detection with Code Language Models: How Far Are We?}. In \bibinfo{booktitle}{\emph{2025 IEEE/ACM 47th International Conference on Software Engineering (ICSE)}}. IEEE Computer Society, \bibinfo{pages}{469--481}.
\newblock


\bibitem[Du et~al\mbox{.}(2024)]%
        {du2024vul}
\bibfield{author}{\bibinfo{person}{Xueying Du}, \bibinfo{person}{Geng Zheng}, \bibinfo{person}{Kaixin Wang}, \bibinfo{person}{Jiayi Feng}, \bibinfo{person}{Wentai Deng}, \bibinfo{person}{Mingwei Liu}, \bibinfo{person}{Bihuan Chen}, \bibinfo{person}{Xin Peng}, \bibinfo{person}{Tao Ma}, {and} \bibinfo{person}{Yiling Lou}.} \bibinfo{year}{2024}\natexlab{}.
\newblock \showarticletitle{Vul-rag: Enhancing llm-based vulnerability detection via knowledge-level rag}.
\newblock \bibinfo{journal}{\emph{arXiv preprint arXiv:2406.11147}} (\bibinfo{year}{2024}).
\newblock


\bibitem[Fan et~al\mbox{.}(2024)]%
        {fan2024survey}
\bibfield{author}{\bibinfo{person}{Wenqi Fan}, \bibinfo{person}{Yujuan Ding}, \bibinfo{person}{Liangbo Ning}, \bibinfo{person}{Shijie Wang}, \bibinfo{person}{Hengyun Li}, \bibinfo{person}{Dawei Yin}, \bibinfo{person}{Tat-Seng Chua}, {and} \bibinfo{person}{Qing Li}.} \bibinfo{year}{2024}\natexlab{}.
\newblock \showarticletitle{A survey on rag meeting llms: Towards retrieval-augmented large language models}. In \bibinfo{booktitle}{\emph{Proceedings of the 30th ACM SIGKDD Conference on Knowledge Discovery and Data Mining}}. \bibinfo{pages}{6491--6501}.
\newblock


\bibitem[Gao et~al\mbox{.}(2023)]%
        {gao2023far}
\bibfield{author}{\bibinfo{person}{Zeyu Gao}, \bibinfo{person}{Hao Wang}, \bibinfo{person}{Yuchen Zhou}, \bibinfo{person}{Wenyu Zhu}, {and} \bibinfo{person}{Chao Zhang}.} \bibinfo{year}{2023}\natexlab{}.
\newblock \showarticletitle{How far have we gone in vulnerability detection using large language models}.
\newblock \bibinfo{journal}{\emph{arXiv preprint arXiv:2311.12420}} (\bibinfo{year}{2023}).
\newblock


\bibitem[GitHub(2025)]%
        {github_copilot}
\bibfield{author}{\bibinfo{person}{GitHub}.} \bibinfo{year}{2025}\natexlab{}.
\newblock \bibinfo{title}{GitHub Copilot · Your AI Pair Programmer}.
\newblock
\newblock
\urldef\tempurl%
\url{https://github.com/features/copilot}
\showURL{%
\tempurl}


\bibitem[Gon{\c{c}}alves et~al\mbox{.}(2024)]%
        {gonccalves2024scope}
\bibfield{author}{\bibinfo{person}{Jos{\'e} Gon{\c{c}}alves}, \bibinfo{person}{Tiago Dias}, \bibinfo{person}{Eva Maia}, {and} \bibinfo{person}{Isabel Pra{\c{c}}a}.} \bibinfo{year}{2024}\natexlab{}.
\newblock \showarticletitle{Scope: Evaluating llms for software vulnerability detection}.
\newblock \bibinfo{journal}{\emph{arXiv preprint arXiv:2407.14372}} (\bibinfo{year}{2024}).
\newblock


\bibitem[Guo et~al\mbox{.}(2025)]%
        {guo2025deepseek}
\bibfield{author}{\bibinfo{person}{Daya Guo}, \bibinfo{person}{Dejian Yang}, \bibinfo{person}{Haowei Zhang}, \bibinfo{person}{Junxiao Song}, \bibinfo{person}{Ruoyu Zhang}, \bibinfo{person}{Runxin Xu}, \bibinfo{person}{Qihao Zhu}, \bibinfo{person}{Shirong Ma}, \bibinfo{person}{Peiyi Wang}, \bibinfo{person}{Xiao Bi}, {et~al\mbox{.}}} \bibinfo{year}{2025}\natexlab{}.
\newblock \showarticletitle{Deepseek-r1: Incentivizing reasoning capability in llms via reinforcement learning}.
\newblock \bibinfo{journal}{\emph{arXiv preprint arXiv:2501.12948}} (\bibinfo{year}{2025}).
\newblock


\bibitem[Havrilla et~al\mbox{.}(2024)]%
        {havrilla2024teaching}
\bibfield{author}{\bibinfo{person}{Alex Havrilla}, \bibinfo{person}{Yuqing Du}, \bibinfo{person}{Sharath~Chandra Raparthy}, \bibinfo{person}{Christoforos Nalmpantis}, \bibinfo{person}{Jane Dwivedi-Yu}, \bibinfo{person}{Maksym Zhuravinskyi}, \bibinfo{person}{Eric Hambro}, \bibinfo{person}{Sainbayar Sukhbaatar}, {and} \bibinfo{person}{Roberta Raileanu}.} \bibinfo{year}{2024}\natexlab{}.
\newblock \showarticletitle{Teaching large language models to reason with reinforcement learning}.
\newblock \bibinfo{journal}{\emph{arXiv preprint arXiv:2403.04642}} (\bibinfo{year}{2024}).
\newblock


\bibitem[Hendrycks et~al\mbox{.}(2021)]%
        {hendrycks2021measuring}
\bibfield{author}{\bibinfo{person}{Dan Hendrycks}, \bibinfo{person}{Collin Burns}, \bibinfo{person}{Saurav Kadavath}, \bibinfo{person}{Akul Arora}, \bibinfo{person}{Steven Basart}, \bibinfo{person}{Eric Tang}, \bibinfo{person}{Dawn Song}, {and} \bibinfo{person}{Jacob Steinhardt}.} \bibinfo{year}{2021}\natexlab{}.
\newblock \showarticletitle{Measuring mathematical problem solving with the math dataset}.
\newblock \bibinfo{journal}{\emph{arXiv preprint arXiv:2103.03874}} (\bibinfo{year}{2021}).
\newblock


\bibitem[Huang and Chang(2022)]%
        {huang2022towards}
\bibfield{author}{\bibinfo{person}{Jie Huang} {and} \bibinfo{person}{Kevin Chen-Chuan Chang}.} \bibinfo{year}{2022}\natexlab{}.
\newblock \showarticletitle{Towards reasoning in large language models: A survey}.
\newblock \bibinfo{journal}{\emph{arXiv preprint arXiv:2212.10403}} (\bibinfo{year}{2022}).
\newblock


\bibitem[IBM(2024)]%
        {cost_of_data_breach_2024_ibm}
\bibfield{author}{\bibinfo{person}{IBM}.} \bibinfo{year}{2024}\natexlab{}.
\newblock \bibinfo{title}{Cost of a Data Breach 2024}.
\newblock
\newblock
\urldef\tempurl%
\url{https://www.ibm.com/reports/data-breach}
\showURL{%
\tempurl}


\bibitem[Ji et~al\mbox{.}(2025)]%
        {ji2025test}
\bibfield{author}{\bibinfo{person}{Yixin Ji}, \bibinfo{person}{Juntao Li}, \bibinfo{person}{Hai Ye}, \bibinfo{person}{Kaixin Wu}, \bibinfo{person}{Jia Xu}, \bibinfo{person}{Linjian Mo}, {and} \bibinfo{person}{Min Zhang}.} \bibinfo{year}{2025}\natexlab{}.
\newblock \showarticletitle{Test-time Computing: from System-1 Thinking to System-2 Thinking}.
\newblock \bibinfo{journal}{\emph{arXiv preprint arXiv:2501.02497}} (\bibinfo{year}{2025}).
\newblock


\bibitem[Joern.io(2024)]%
        {joern_bug_hunters_workbench_2024}
\bibfield{author}{\bibinfo{person}{Joern.io}.} \bibinfo{year}{2024}\natexlab{}.
\newblock \bibinfo{title}{Joern - The Bug Hunter’s Workbench}.
\newblock
\newblock
\urldef\tempurl%
\url{https://joern.io/}
\showURL{%
\tempurl}


\bibitem[Kahneman(2011)]%
        {kahneman2011thinking}
\bibfield{author}{\bibinfo{person}{Daniel Kahneman}.} \bibinfo{year}{2011}\natexlab{}.
\newblock \bibinfo{booktitle}{\emph{Thinking, fast and slow}}.
\newblock \bibinfo{publisher}{macmillan}.
\newblock


\bibitem[Kaplan et~al\mbox{.}(2020)]%
        {kaplan2020scaling}
\bibfield{author}{\bibinfo{person}{Jared Kaplan}, \bibinfo{person}{Sam McCandlish}, \bibinfo{person}{Tom Henighan}, \bibinfo{person}{Tom~B Brown}, \bibinfo{person}{Benjamin Chess}, \bibinfo{person}{Rewon Child}, \bibinfo{person}{Scott Gray}, \bibinfo{person}{Alec Radford}, \bibinfo{person}{Jeffrey Wu}, {and} \bibinfo{person}{Dario Amodei}.} \bibinfo{year}{2020}\natexlab{}.
\newblock \showarticletitle{Scaling laws for neural language models}.
\newblock \bibinfo{journal}{\emph{arXiv preprint arXiv:2001.08361}} (\bibinfo{year}{2020}).
\newblock


\bibitem[Khare et~al\mbox{.}(2023)]%
        {khare2023understanding}
\bibfield{author}{\bibinfo{person}{Avishree Khare}, \bibinfo{person}{Saikat Dutta}, \bibinfo{person}{Ziyang Li}, \bibinfo{person}{Alaia Solko-Breslin}, \bibinfo{person}{Rajeev Alur}, {and} \bibinfo{person}{Mayur Naik}.} \bibinfo{year}{2023}\natexlab{}.
\newblock \showarticletitle{Understanding the effectiveness of large language models in detecting security vulnerabilities}.
\newblock \bibinfo{journal}{\emph{arXiv preprint arXiv:2311.16169}} (\bibinfo{year}{2023}).
\newblock


\bibitem[Lewis et~al\mbox{.}(2020)]%
        {lewis2020retrieval}
\bibfield{author}{\bibinfo{person}{Patrick Lewis}, \bibinfo{person}{Ethan Perez}, \bibinfo{person}{Aleksandra Piktus}, \bibinfo{person}{Fabio Petroni}, \bibinfo{person}{Vladimir Karpukhin}, \bibinfo{person}{Naman Goyal}, \bibinfo{person}{Heinrich K{\"u}ttler}, \bibinfo{person}{Mike Lewis}, \bibinfo{person}{Wen-tau Yih}, \bibinfo{person}{Tim Rockt{\"a}schel}, {et~al\mbox{.}}} \bibinfo{year}{2020}\natexlab{}.
\newblock \showarticletitle{Retrieval-augmented generation for knowledge-intensive nlp tasks}.
\newblock \bibinfo{journal}{\emph{Advances in neural information processing systems}}  \bibinfo{volume}{33} (\bibinfo{year}{2020}), \bibinfo{pages}{9459--9474}.
\newblock


\bibitem[Lightman et~al\mbox{.}(2023)]%
        {lightman2023let}
\bibfield{author}{\bibinfo{person}{Hunter Lightman}, \bibinfo{person}{Vineet Kosaraju}, \bibinfo{person}{Yuri Burda}, \bibinfo{person}{Harrison Edwards}, \bibinfo{person}{Bowen Baker}, \bibinfo{person}{Teddy Lee}, \bibinfo{person}{Jan Leike}, \bibinfo{person}{John Schulman}, \bibinfo{person}{Ilya Sutskever}, {and} \bibinfo{person}{Karl Cobbe}.} \bibinfo{year}{2023}\natexlab{}.
\newblock \showarticletitle{Let's verify step by step}. In \bibinfo{booktitle}{\emph{The Twelfth International Conference on Learning Representations}}.
\newblock


\bibitem[Liu et~al\mbox{.}(2024)]%
        {liu2024vuldetectbench}
\bibfield{author}{\bibinfo{person}{Yu Liu}, \bibinfo{person}{Lang Gao}, \bibinfo{person}{Mingxin Yang}, \bibinfo{person}{Yu Xie}, \bibinfo{person}{Ping Chen}, \bibinfo{person}{Xiaojin Zhang}, {and} \bibinfo{person}{Wei Chen}.} \bibinfo{year}{2024}\natexlab{}.
\newblock \showarticletitle{Vuldetectbench: Evaluating the deep capability of vulnerability detection with large language models}.
\newblock \bibinfo{journal}{\emph{arXiv preprint arXiv:2406.07595}} (\bibinfo{year}{2024}).
\newblock


\bibitem[Ma et~al\mbox{.}(2024)]%
        {ma2024combining}
\bibfield{author}{\bibinfo{person}{Wei Ma}, \bibinfo{person}{Daoyuan Wu}, \bibinfo{person}{Yuqiang Sun}, \bibinfo{person}{Tianwen Wang}, \bibinfo{person}{Shangqing Liu}, \bibinfo{person}{Jian Zhang}, \bibinfo{person}{Yue Xue}, {and} \bibinfo{person}{Yang Liu}.} \bibinfo{year}{2024}\natexlab{}.
\newblock \showarticletitle{Combining fine-tuning and llm-based agents for intuitive smart contract auditing with justifications}.
\newblock \bibinfo{journal}{\emph{arXiv preprint arXiv:2403.16073}} (\bibinfo{year}{2024}).
\newblock


\bibitem[Menguy et~al\mbox{.}(2022)]%
        {menguy2022automated}
\bibfield{author}{\bibinfo{person}{Gr{\'e}goire Menguy}, \bibinfo{person}{S{\'e}bastien Bardin}, \bibinfo{person}{Nadjib Lazaar}, {and} \bibinfo{person}{Arnaud Gotlieb}.} \bibinfo{year}{2022}\natexlab{}.
\newblock \showarticletitle{Automated program analysis: Revisiting precondition inference through constraint acquisition}. In \bibinfo{booktitle}{\emph{IJCAI-ECAI 22-31st International Joint Conference on Artificial Intelligence and the 25th European Conference on Artificial Intelligence}}. Curran Associates, Inc.(May 2023), \bibinfo{pages}{1873--1879}.
\newblock


\bibitem[Muennighoff et~al\mbox{.}(2025)]%
        {muennighoff2025s1}
\bibfield{author}{\bibinfo{person}{Niklas Muennighoff}, \bibinfo{person}{Zitong Yang}, \bibinfo{person}{Weijia Shi}, \bibinfo{person}{Xiang~Lisa Li}, \bibinfo{person}{Li Fei-Fei}, \bibinfo{person}{Hannaneh Hajishirzi}, \bibinfo{person}{Luke Zettlemoyer}, \bibinfo{person}{Percy Liang}, \bibinfo{person}{Emmanuel Cand{\`e}s}, {and} \bibinfo{person}{Tatsunori Hashimoto}.} \bibinfo{year}{2025}\natexlab{}.
\newblock \showarticletitle{s1: Simple test-time scaling}.
\newblock \bibinfo{journal}{\emph{arXiv preprint arXiv:2501.19393}} (\bibinfo{year}{2025}).
\newblock


\bibitem[Nong et~al\mbox{.}(2024)]%
        {nong2024chain}
\bibfield{author}{\bibinfo{person}{Yu Nong}, \bibinfo{person}{Mohammed Aldeen}, \bibinfo{person}{Long Cheng}, \bibinfo{person}{Hongxin Hu}, \bibinfo{person}{Feng Chen}, {and} \bibinfo{person}{Haipeng Cai}.} \bibinfo{year}{2024}\natexlab{}.
\newblock \showarticletitle{Chain-of-thought prompting of large language models for discovering and fixing software vulnerabilities}.
\newblock \bibinfo{journal}{\emph{arXiv preprint arXiv:2402.17230}} (\bibinfo{year}{2024}).
\newblock


\bibitem[of~Standards and (NIST)(2025)]%
        {NIST-NVD-2019}
\bibfield{author}{\bibinfo{person}{National~Institute of Standards} {and} \bibinfo{person}{Technology (NIST)}.} \bibinfo{year}{2025}\natexlab{}.
\newblock \bibinfo{title}{National Vulnerability Database (NVD)}.
\newblock
\newblock
\urldef\tempurl%
\url{https://nvd.nist.gov/}
\showURL{%
\tempurl}


\bibitem[OpenAI(2025)]%
        {openai_o3_mini_system_card_2025}
\bibfield{author}{\bibinfo{person}{OpenAI}.} \bibinfo{year}{2025}\natexlab{}.
\newblock \bibinfo{title}{o3-mini System Card}.
\newblock
\newblock
\urldef\tempurl%
\url{https://openai.com/index/o3-mini-system-card/}
\showURL{%
\tempurl}


\bibitem[Padhi et~al\mbox{.}(2016)]%
        {padhi2016data}
\bibfield{author}{\bibinfo{person}{Saswat Padhi}, \bibinfo{person}{Rahul Sharma}, {and} \bibinfo{person}{Todd Millstein}.} \bibinfo{year}{2016}\natexlab{}.
\newblock \showarticletitle{Data-driven precondition inference with learned features}.
\newblock \bibinfo{journal}{\emph{ACM SIGPLAN Notices}} \bibinfo{volume}{51}, \bibinfo{number}{6} (\bibinfo{year}{2016}), \bibinfo{pages}{42--56}.
\newblock


\bibitem[Pan et~al\mbox{.}(2024)]%
        {pan2024automatically}
\bibfield{author}{\bibinfo{person}{Liangming Pan}, \bibinfo{person}{Michael Saxon}, \bibinfo{person}{Wenda Xu}, \bibinfo{person}{Deepak Nathani}, \bibinfo{person}{Xinyi Wang}, {and} \bibinfo{person}{William~Yang Wang}.} \bibinfo{year}{2024}\natexlab{}.
\newblock \showarticletitle{Automatically correcting large language models: Surveying the landscape of diverse automated correction strategies}.
\newblock \bibinfo{journal}{\emph{Transactions of the Association for Computational Linguistics}}  \bibinfo{volume}{12} (\bibinfo{year}{2024}), \bibinfo{pages}{484--506}.
\newblock


\bibitem[Project(2021)]%
        {gnu_cflow_2021}
\bibfield{author}{\bibinfo{person}{GNU Project}.} \bibinfo{year}{2021}\natexlab{}.
\newblock \bibinfo{title}{GNU Cflow - Free Software Foundation}.
\newblock
\newblock
\urldef\tempurl%
\url{https://www.gnu.org/software/cflow/}
\showURL{%
\tempurl}


\bibitem[Rajani et~al\mbox{.}(2019)]%
        {rajani2019explain}
\bibfield{author}{\bibinfo{person}{Nazneen~Fatema Rajani}, \bibinfo{person}{Bryan McCann}, \bibinfo{person}{Caiming Xiong}, {and} \bibinfo{person}{Richard Socher}.} \bibinfo{year}{2019}\natexlab{}.
\newblock \showarticletitle{Explain yourself! leveraging language models for commonsense reasoning}.
\newblock \bibinfo{journal}{\emph{arXiv preprint arXiv:1906.02361}} (\bibinfo{year}{2019}).
\newblock


\bibitem[Risse and B{\"o}hme(2024a)]%
        {risse2024top}
\bibfield{author}{\bibinfo{person}{Niklas Risse} {and} \bibinfo{person}{Marcel B{\"o}hme}.} \bibinfo{year}{2024}\natexlab{a}.
\newblock \showarticletitle{Top score on the wrong exam: On benchmarking in machine learning for vulnerability detection}.
\newblock \bibinfo{journal}{\emph{arXiv preprint arXiv:2408.12986}} (\bibinfo{year}{2024}).
\newblock


\bibitem[Risse and B{\"o}hme(2024b)]%
        {risse2024uncovering}
\bibfield{author}{\bibinfo{person}{Niklas Risse} {and} \bibinfo{person}{Marcel B{\"o}hme}.} \bibinfo{year}{2024}\natexlab{b}.
\newblock \showarticletitle{Uncovering the limits of machine learning for automatic vulnerability detection}. In \bibinfo{booktitle}{\emph{33rd USENIX Security Symposium (USENIX Security 24)}}. \bibinfo{pages}{4247--4264}.
\newblock


\bibitem[Ristea et~al\mbox{.}(2024)]%
        {ristea2024benchmarking}
\bibfield{author}{\bibinfo{person}{Dan Ristea}, \bibinfo{person}{Vasilios Mavroudis}, {and} \bibinfo{person}{Chris Hicks}.} \bibinfo{year}{2024}\natexlab{}.
\newblock \showarticletitle{Benchmarking OpenAI o1 in Cyber Security}.
\newblock \bibinfo{journal}{\emph{arXiv preprint arXiv:2410.21939}} (\bibinfo{year}{2024}).
\newblock


\bibitem[Sheng et~al\mbox{.}(2025a)]%
        {sheng2025large}
\bibfield{author}{\bibinfo{person}{Ze Sheng}, \bibinfo{person}{Zhicheng Chen}, \bibinfo{person}{Shuning Gu}, \bibinfo{person}{Heqing Huang}, \bibinfo{person}{Guofei Gu}, {and} \bibinfo{person}{Jeff Huang}.} \bibinfo{year}{2025}\natexlab{a}.
\newblock \showarticletitle{Large Language Models in Software Security: A Survey of Vulnerability Detection Techniques and Insights}.
\newblock \bibinfo{journal}{\emph{arXiv preprint arXiv:2502.07049}} (\bibinfo{year}{2025}).
\newblock


\bibitem[Sheng et~al\mbox{.}(2025b)]%
        {sheng2025llms}
\bibfield{author}{\bibinfo{person}{Ze Sheng}, \bibinfo{person}{Zhicheng Chen}, \bibinfo{person}{Shuning Gu}, \bibinfo{person}{Heqing Huang}, \bibinfo{person}{Guofei Gu}, {and} \bibinfo{person}{Jeff Huang}.} \bibinfo{year}{2025}\natexlab{b}.
\newblock \showarticletitle{LLMs in Software Security: A Survey of Vulnerability Detection Techniques and Insights}.
\newblock \bibinfo{journal}{\emph{arXiv e-prints}} (\bibinfo{year}{2025}), \bibinfo{pages}{arXiv--2502}.
\newblock


\bibitem[Sloman(1996)]%
        {sloman1996empirical}
\bibfield{author}{\bibinfo{person}{Steven~A Sloman}.} \bibinfo{year}{1996}\natexlab{}.
\newblock \showarticletitle{The empirical case for two systems of reasoning.}
\newblock \bibinfo{journal}{\emph{Psychological bulletin}} \bibinfo{volume}{119}, \bibinfo{number}{1} (\bibinfo{year}{1996}), \bibinfo{pages}{3}.
\newblock


\bibitem[Snell et~al\mbox{.}(2024)]%
        {snell2024scaling}
\bibfield{author}{\bibinfo{person}{Charlie Snell}, \bibinfo{person}{Jaehoon Lee}, \bibinfo{person}{Kelvin Xu}, {and} \bibinfo{person}{Aviral Kumar}.} \bibinfo{year}{2024}\natexlab{}.
\newblock \showarticletitle{Scaling llm test-time compute optimally can be more effective than scaling model parameters}.
\newblock \bibinfo{journal}{\emph{arXiv preprint arXiv:2408.03314}} (\bibinfo{year}{2024}).
\newblock


\bibitem[Steenhoek et~al\mbox{.}(2024)]%
        {steenhoek2024err}
\bibfield{author}{\bibinfo{person}{Benjamin Steenhoek}, \bibinfo{person}{Md~Mahbubur Rahman}, \bibinfo{person}{Monoshi~Kumar Roy}, \bibinfo{person}{Mirza~Sanjida Alam}, \bibinfo{person}{Hengbo Tong}, \bibinfo{person}{Swarna Das}, \bibinfo{person}{Earl~T Barr}, {and} \bibinfo{person}{Wei Le}.} \bibinfo{year}{2024}\natexlab{}.
\newblock \showarticletitle{To Err is Machine: Vulnerability Detection Challenges LLM Reasoning}.
\newblock \bibinfo{journal}{\emph{arXiv preprint arXiv:2403.17218}} (\bibinfo{year}{2024}).
\newblock


\bibitem[Stenning and Van~Lambalgen(2012)]%
        {stenning2012human}
\bibfield{author}{\bibinfo{person}{Keith Stenning} {and} \bibinfo{person}{Michiel Van~Lambalgen}.} \bibinfo{year}{2012}\natexlab{}.
\newblock \bibinfo{booktitle}{\emph{Human reasoning and cognitive science}}.
\newblock \bibinfo{publisher}{MIT Press}.
\newblock


\bibitem[Sun et~al\mbox{.}(2024)]%
        {sun2024llm4vuln}
\bibfield{author}{\bibinfo{person}{Yuqiang Sun}, \bibinfo{person}{Daoyuan Wu}, \bibinfo{person}{Yue Xue}, \bibinfo{person}{Han Liu}, \bibinfo{person}{Wei Ma}, \bibinfo{person}{Lyuye Zhang}, \bibinfo{person}{Yang Liu}, {and} \bibinfo{person}{Yingjiu Li}.} \bibinfo{year}{2024}\natexlab{}.
\newblock \showarticletitle{Llm4vuln: A unified evaluation framework for decoupling and enhancing llms' vulnerability reasoning}.
\newblock \bibinfo{journal}{\emph{arXiv preprint arXiv:2401.16185}} (\bibinfo{year}{2024}).
\newblock


\bibitem[Tan et~al\mbox{.}(2024)]%
        {tan2024judgebench}
\bibfield{author}{\bibinfo{person}{Sijun Tan}, \bibinfo{person}{Siyuan Zhuang}, \bibinfo{person}{Kyle Montgomery}, \bibinfo{person}{William~Y Tang}, \bibinfo{person}{Alejandro Cuadron}, \bibinfo{person}{Chenguang Wang}, \bibinfo{person}{Raluca~Ada Popa}, {and} \bibinfo{person}{Ion Stoica}.} \bibinfo{year}{2024}\natexlab{}.
\newblock \showarticletitle{Judgebench: A benchmark for evaluating llm-based judges}.
\newblock \bibinfo{journal}{\emph{arXiv preprint arXiv:2410.12784}} (\bibinfo{year}{2024}).
\newblock


\bibitem[Ullah et~al\mbox{.}(2024)]%
        {ullah2024llms}
\bibfield{author}{\bibinfo{person}{Saad Ullah}, \bibinfo{person}{Mingji Han}, \bibinfo{person}{Saurabh Pujar}, \bibinfo{person}{Hammond Pearce}, \bibinfo{person}{Ayse Coskun}, {and} \bibinfo{person}{Gianluca Stringhini}.} \bibinfo{year}{2024}\natexlab{}.
\newblock \showarticletitle{Llms cannot reliably identify and reason about security vulnerabilities (yet?): A comprehensive evaluation, framework, and benchmarks}. In \bibinfo{booktitle}{\emph{2024 IEEE Symposium on Security and Privacy (SP)}}. IEEE, \bibinfo{pages}{862--880}.
\newblock


\bibitem[Wang et~al\mbox{.}(2024b)]%
        {wang2024advanced}
\bibfield{author}{\bibinfo{person}{Guoqing Wang}, \bibinfo{person}{Zeyu Sun}, \bibinfo{person}{Zhihao Gong}, \bibinfo{person}{Sixiang Ye}, \bibinfo{person}{Yizhou Chen}, \bibinfo{person}{Yifan Zhao}, \bibinfo{person}{Qingyuan Liang}, {and} \bibinfo{person}{Dan Hao}.} \bibinfo{year}{2024}\natexlab{b}.
\newblock \showarticletitle{Do advanced language models eliminate the need for prompt engineering in software engineering?}
\newblock \bibinfo{journal}{\emph{arXiv preprint arXiv:2411.02093}} (\bibinfo{year}{2024}).
\newblock


\bibitem[Wang et~al\mbox{.}(2024a)]%
        {wang2024reposvul}
\bibfield{author}{\bibinfo{person}{Xinchen Wang}, \bibinfo{person}{Ruida Hu}, \bibinfo{person}{Cuiyun Gao}, \bibinfo{person}{Xin-Cheng Wen}, \bibinfo{person}{Yujia Chen}, {and} \bibinfo{person}{Qing Liao}.} \bibinfo{year}{2024}\natexlab{a}.
\newblock \showarticletitle{Reposvul: A repository-level high-quality vulnerability dataset}. In \bibinfo{booktitle}{\emph{Proceedings of the 2024 IEEE/ACM 46th International Conference on Software Engineering: Companion Proceedings}}. \bibinfo{pages}{472--483}.
\newblock


\bibitem[Wei et~al\mbox{.}(2022)]%
        {wei2022chain}
\bibfield{author}{\bibinfo{person}{Jason Wei}, \bibinfo{person}{Xuezhi Wang}, \bibinfo{person}{Dale Schuurmans}, \bibinfo{person}{Maarten Bosma}, \bibinfo{person}{Fei Xia}, \bibinfo{person}{Ed Chi}, \bibinfo{person}{Quoc~V Le}, \bibinfo{person}{Denny Zhou}, {et~al\mbox{.}}} \bibinfo{year}{2022}\natexlab{}.
\newblock \showarticletitle{Chain-of-thought prompting elicits reasoning in large language models}.
\newblock \bibinfo{journal}{\emph{Advances in neural information processing systems}}  \bibinfo{volume}{35} (\bibinfo{year}{2022}), \bibinfo{pages}{24824--24837}.
\newblock


\bibitem[Wen et~al\mbox{.}(2024)]%
        {wen2024vuleval}
\bibfield{author}{\bibinfo{person}{Xin-Cheng Wen}, \bibinfo{person}{Xinchen Wang}, \bibinfo{person}{Yujia Chen}, \bibinfo{person}{Ruida Hu}, \bibinfo{person}{David Lo}, {and} \bibinfo{person}{Cuiyun Gao}.} \bibinfo{year}{2024}\natexlab{}.
\newblock \showarticletitle{Vuleval: Towards repository-level evaluation of software vulnerability detection}.
\newblock \bibinfo{journal}{\emph{arXiv preprint arXiv:2404.15596}} (\bibinfo{year}{2024}).
\newblock


\bibitem[Yao et~al\mbox{.}(2023)]%
        {yao2023tree}
\bibfield{author}{\bibinfo{person}{Shunyu Yao}, \bibinfo{person}{Dian Yu}, \bibinfo{person}{Jeffrey Zhao}, \bibinfo{person}{Izhak Shafran}, \bibinfo{person}{Tom Griffiths}, \bibinfo{person}{Yuan Cao}, {and} \bibinfo{person}{Karthik Narasimhan}.} \bibinfo{year}{2023}\natexlab{}.
\newblock \showarticletitle{Tree of thoughts: Deliberate problem solving with large language models}.
\newblock \bibinfo{journal}{\emph{Advances in neural information processing systems}}  \bibinfo{volume}{36} (\bibinfo{year}{2023}), \bibinfo{pages}{11809--11822}.
\newblock


\bibitem[Yildiz et~al\mbox{.}(2025)]%
        {yildiz2025benchmarking}
\bibfield{author}{\bibinfo{person}{Alperen Yildiz}, \bibinfo{person}{Sin~G Teo}, \bibinfo{person}{Yiling Lou}, \bibinfo{person}{Yebo Feng}, \bibinfo{person}{Chong Wang}, {and} \bibinfo{person}{Dinil~M Divakaran}.} \bibinfo{year}{2025}\natexlab{}.
\newblock \showarticletitle{Benchmarking LLMs and LLM-based Agents in Practical Vulnerability Detection for Code Repositories}.
\newblock \bibinfo{journal}{\emph{arXiv preprint arXiv:2503.03586}} (\bibinfo{year}{2025}).
\newblock


\bibitem[Yin et~al\mbox{.}(2024)]%
        {yin2024multitask}
\bibfield{author}{\bibinfo{person}{Xin Yin}, \bibinfo{person}{Chao Ni}, {and} \bibinfo{person}{Shaohua Wang}.} \bibinfo{year}{2024}\natexlab{}.
\newblock \showarticletitle{Multitask-based evaluation of open-source llm on software vulnerability}.
\newblock \bibinfo{journal}{\emph{IEEE Transactions on Software Engineering}} (\bibinfo{year}{2024}).
\newblock


\bibitem[Yu et~al\mbox{.}(2024)]%
        {yu2024distilling}
\bibfield{author}{\bibinfo{person}{Ping Yu}, \bibinfo{person}{Jing Xu}, \bibinfo{person}{Jason Weston}, {and} \bibinfo{person}{Ilia Kulikov}.} \bibinfo{year}{2024}\natexlab{}.
\newblock \showarticletitle{Distilling system 2 into system 1}.
\newblock \bibinfo{journal}{\emph{arXiv preprint arXiv:2407.06023}} (\bibinfo{year}{2024}).
\newblock


\bibitem[Zibaeirad and Vieira(2024)]%
        {zibaeirad2024vulnllmeval}
\bibfield{author}{\bibinfo{person}{Arastoo Zibaeirad} {and} \bibinfo{person}{Marco Vieira}.} \bibinfo{year}{2024}\natexlab{}.
\newblock \showarticletitle{VulnLLMEval: A Framework for Evaluating Large Language Models in Software Vulnerability Detection and Patching}.
\newblock \bibinfo{journal}{\emph{arXiv preprint arXiv:2409.10756}} (\bibinfo{year}{2024}).
\newblock


\bibitem[Zibaeirad and Vieira(2025)]%
        {zibaeirad2025reasoning}
\bibfield{author}{\bibinfo{person}{Arastoo Zibaeirad} {and} \bibinfo{person}{Marco Vieira}.} \bibinfo{year}{2025}\natexlab{}.
\newblock \showarticletitle{Reasoning with LLMs for Zero-Shot Vulnerability Detection}.
\newblock \bibinfo{journal}{\emph{arXiv preprint arXiv:2503.17885}} (\bibinfo{year}{2025}).
\newblock


\end{thebibliography}

\newpage
\appendix
\section{A RAG Example}
\label{appendix:rag}
\begin{figure}[H]
    \centering
    \includegraphics[width=0.48\textwidth]{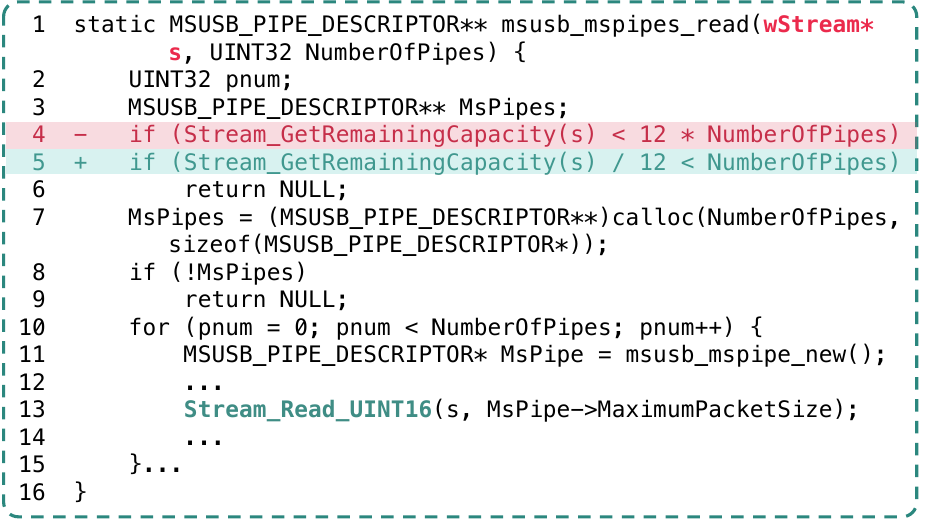}
    \caption{CVE-2020-11039, an integer overflow vulnerability.}
    \label{fig:rag_example}
\end{figure}
\autoref{fig:rag_example} presents one of the patches retrieved using \autoref{fig:code}(a) as the query, based on functional semantic similarity~\cite{du2024vul}. This function reads pipe descriptors from a stream and returns an array of pointers to these descriptors, ensuring correct memory allocation and robust error handling. The operation that reads data from the stream and assigns it to a variable (line 13) closely resembles the corresponding operation in \autoref{fig:code}(a). However, the function contains a vulnerability in line 4, where the computation \textsf{12 * NumberOfPipes} could result in an integer overflow. To mitigate this issue, the patch modifies the operation by replacing the multiplication with a division.

 RAG has two key limitations to consider. First, the similar code retrieved through RAG fails to address the challenge of missing context. Second, since the specific type of vulnerability in the target code is unknown before detection, the patches suggested by RAG may not always enhance the accuracy of vulnerability detection. For example, when \autoref{fig:rag_example} is used as a one-shot example to guide GPT-4o in analyzing \autoref{fig:code}(a), the model might incorrectly identify \textsf{msg\_data} as the vulnerable component, similar to what is illustrated in \autoref{fig:code}(d). This misidentification arises because both \textsf{msg\_data} and line 4 in \autoref{fig:rag_example} involve input validation logic, leading the model to draw parallels that may not be relevant.

\section{CWE-1000: Overview and Classification}
\label{appendix:cwe1000}

The CWE-1000 provides a hierarchical structure for categorizing software weaknesses at various levels of abstraction. At the highest level, these weaknesses are grouped into pillars, which represent the most abstract categories of vulnerabilities. Below is an overview of the key pillars:

\begin{itemize}[partopsep=2pt, topsep=-\parskip, parsep=2pt, itemsep=2pt, leftmargin=*]
    \item \textbf{Improper Access Control}: Weaknesses related to insufficient restrictions on access to resources or functionalities.
    
    \item \textbf{Improper Interaction Between Multiple Correctly-Behaving Entities}: Issues arising from unintended interactions between components that function correctly in isolation.

    \item \textbf{Improper Control of a Resource Through its Lifetime}: Weaknesses involving inadequate management of resources during their creation, usage, or destruction.

    \item \textbf{Incorrect Calculation}: Errors in computations or logical operations that lead to incorrect results.

    \item \textbf{Insufficient Control Flow Management}: Weaknesses related to improper handling of program execution paths.

    \item \textbf{Protection Mechanism Failure}: Failures in security mechanisms designed to protect systems or data.

    \item \textbf{Incorrect Comparison}: Issues caused by flawed comparisons, such as incorrect equality checks.

    \item \textbf{Improper Check or Handling of Exceptional Conditions}: Weaknesses stemming from inadequate handling of error conditions or exceptions.

    \item \textbf{Improper Neutralization}: Vulnerabilities involving failure to sanitize inputs or outputs, potentially leading to injection attacks.

    \item \textbf{Improper Adherence to Coding Standards}: Weaknesses resulting from violations of established coding practices or guidelines.
\end{itemize}

The CWE-1000 is particularly valuable for researchers seeking to identify gaps in knowledge by examining high-level categories with few documented weaknesses. This hierarchical classification enables systematic analysis and understanding of vulnerabilities across different abstraction levels.

To evaluate our methodology, we sampled a subset of 400 data from the full 2,000 entries, adhering to the classification standards outlined in CWE-1000. This sampling process ensures a representative distribution of vulnerabilities across the defined pillars, enabling comprehensive evaluation and analysis.

\section{Feedback Settings in Strict Mode}
\label{appendix:feedback}

\begin{table}[h!]
\centering
\setlength{\tabcolsep}{4pt}
\caption{Feedback Rounds and Case Modifications.}
\footnotesize
\renewcommand{\arraystretch}{1.1}
\begin{tabular}{lccccc}
\toprule
\textbf{Round} & 0 (Init) & 1 & 2 & 3 & 4 \\ 
\midrule
\textbf{Count} & {\small 4647 (100\%)} & {\small 477 (10.3\%)} & {\small 165 (3.6\%)} & {\small 101 (2.2\%)} & {\small 69 (1.5\%)} \\ 
\bottomrule
\end{tabular}
\label{tab:feedback_cases}
\end{table}

We set \textsf{max\_feedback\_rounds} to 4, which is sufficient for evaluation. 
\autoref{tab:feedback_cases} shows the total number of cases (4647) where the 
output is \textbf{NO\_VUL} across all models, as well as the number and percentage 
of cases successfully altered by the feedback mechanism at round $k$ 
($k \in \{1, 2, 3, 4\}$). When $k \geq 2$, the proportion of cases modified by 
the feedback mechanism drops below 5\%. To balance the effectiveness of feedback 
with its computational cost, we choose \textsf{max\_feedback\_rounds} = 4.

\section{Details of Collected Dataset}
\label{appendix:dataset}
This section provides additional information about the dataset collected by \framework.

\begin{figure}[H]
    \centering
    
    % Subfigure 1: Context Length Distribution
    \begin{subfigure}[b]{0.4\textwidth}
        \centering
        \includegraphics[width=\textwidth]{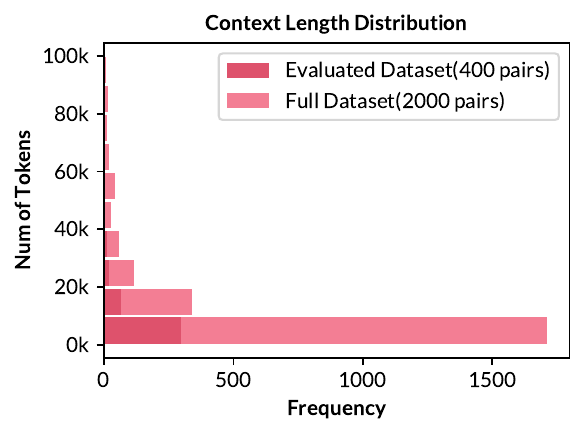}
        \caption{Context length distribution of the evaluated dataset and full dataset.}
        \label{fig:context_length}
    \end{subfigure}
    \hfill
    
    % Subfigure 2: Function Length Distribution
    \begin{subfigure}[b]{0.39\textwidth}
        \hspace{0.5mm}
        % \centering
        \includegraphics[width=\textwidth]{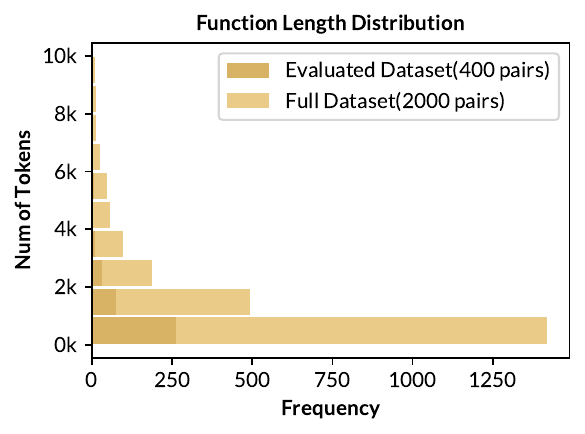}
        \caption{Function length distribution of the evaluated dataset and full dataset.}
        \label{fig:function_length}
    \end{subfigure}
    \hfill
    
    % Subfigure 3: Top 15 Projects Distribution
    \begin{subfigure}[b]{0.405\textwidth}
        \hspace{-1.5mm}
        \includegraphics[width=\textwidth]{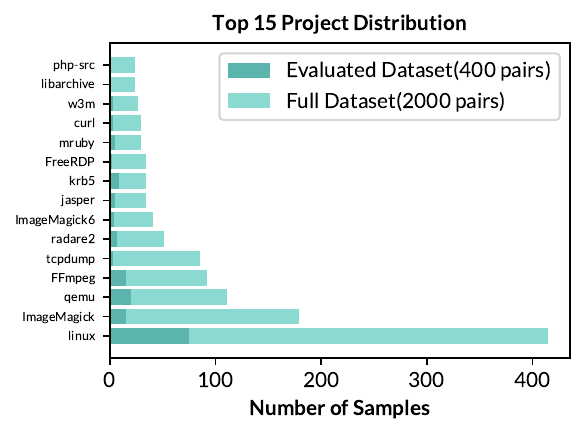}
        \caption{Top 15 projects distribution of the evaluated dataset and full dataset.}
        \label{fig:projects}
    \end{subfigure}
    
    \caption{Distributions of context length, function length, and top 15 projects.}
    \label{fig:combined_distributions}
\end{figure}

\autoref{fig:context_length}, \autoref{fig:function_length}, and \autoref{fig:projects} respectively illustrate the distributions of context length, vulnerable (patched) function length, and the source projects in all collected data and the evaluated data. As can be observed, the majority of contexts are concentrated in the range of 0–20k tokens, while the lengths of vulnerable (patched) functions are predominantly within 0–2k tokens. Additionally, the data originates from a variety of real-world projects, showcasing its diversity and practical relevance.

\section{Prompts Used in \framework}
\label{appendix:prompt}
\autoref{pmt:detection} presents the Context-Rich Vulnerability Assessment Prompt, which is used to generate the detection output of the evaluated LLM, including the rationale and the conclusion.
\begin{figure}[H]
\begin{mybox}{\textbf{\textit{Context-Rich Vulnerability Assessment Prompt}}}
Your task is to evaluate whether the following code contains any of the following vulnerabilities.\\
   \textbf{CWE description}: \\ \textcolor{gray}{\textit{\# including the ground-truth CWE and its description.}} \\
   \textbf{Context-Rich Code}: \\ \textcolor{gray}{\textit{\# including context and the code to be detected, the slicing path of vulnerabilities in the code will be marked with //potential. }} \\
   \textbf{Assumptions}:\\ \textcolor{gray}{\textit{\# direct the LLM to concentrate only on //potential and vuln-related parameters. }} \\
   Analyze the code step by step to determine if any of the specified vulnerabilities are present. In your final response, list all detected vulnerabilities and indicate "HAS\_VUL" if any are found. If no vulnerabilities are detected, respond with "NO\_VUL".
\end{mybox}
\caption{Context-rich vulnerability assessment prompt.}
\label{pmt:detection}
\end{figure}

\begin{figure}[H]
\begin{mybox}{\textbf{\textit{Rationale Assessment Prompt (for vulnerable input)}}}
You are a security expert tasked with evaluating a vulnerability detection tool. You are provided with the following:\\
* Ground Truth: This includes a CVE description, a CWE ID, a commit, and a commit message, which collectively describe the cause of the vulnerability.\\
* Rationale: This is a vulnerability detection rationale generated by a tool, explaining the detected causes of the vulnerability.\\
   \textbf{CVE description}: \\
   \textbf{CWE description}: \\
   \textbf{Commit message}:\\
   \textbf{Commit diff (line by line)}:\\ \textcolor{gray}{\textit{\# a diff view of the commit, including the original code along with the deleted and added lines.}} \\
   \textbf{Rationale}:\\
   The rationale is generated based on the vulnerable version of the code, rather than the patched code. This does not necessarily mean the vulnerability detection tool has produced a correct result. We are specifically interested in whether the rationale correctly identifies the ground-truth vulnerability.\\
If the causes described in the rationale include the ground-truth vulnerability, even if it also mentions unrelated issues, it indicates a MATCH.\\
If the rationale does not include the ground-truth vulnerability and only identifies unrelated issues, return MISMATCH.\\
Let's think step by step, first analyze the ground-truth and rationale, in the end return "MATCH" or "MISMATCH".
\end{mybox}
\caption{Rationale assessment prompt (for vulnerable input).}
\label{pmt:assessmentvuln}
\end{figure}

\autoref{pmt:assessmentvuln} and \autoref{pmt:assessmentpatched} respectively present the prompts used for LLM-as-a-judge evaluation rationales. The two prompts share the same structure but differ in their instructions. The purpose is to assess whether the rationale correctly identifies the ground-truth vulnerabilities or falsely considers patched vulnerabilities as still vulnerable.

\begin{figure}[H]
\begin{mybox}{\textbf{\textit{Rationale Assessment Prompt (for patched input)}}}
You are a security expert tasked with evaluating a vulnerability detection tool. You are provided with the following:\\
* Ground Truth: This includes a CVE description, a CWE ID, a commit, and a commit message, which collectively describe the cause of the vulnerability.\\
* Rationale: This is a vulnerability detection rationale generated by a tool, explaining the detected causes of the vulnerability.\\
   \textbf{CVE description}: \\
   \textbf{CWE description}: \\
   \textbf{Commit message}:\\
   \textbf{Commit diff (line by line)}:\\
   \textbf{Rationale}:\\
   The rationale is generated based on the patched version of the code, not the original vulnerable code, which means the tool reports some issues on the non-vulnerable code. However, this does not necessarily mean the vulnerability detection tool has produced a false alarm. We are specifically interested in whether the rationale includes a false alarm related to the ground-truth vulnerability.\\
If the causes described in the rationale include the ground-truth vulnerability (already fixed in the patched code), meaning either the rationale considers a newly added line in the patch problematic (indicated by + in the diff), or the cause identified by the rationale matches the ground-truth vulnerability, it indicates a FALSE\_ALARM.\\
Otherwise, if the rationale does not include the ground-truth vulnerability or refers to different issues, return CORRECT.\\
Let's think step by step, first analyze the ground-truth and rationale, in the end return "FALSE\_ALARM" or "CORRECT".
\end{mybox}
\caption{Rationale assessment prompt (for patched input).}
\label{pmt:assessmentpatched}
\end{figure}

\section{Further Details About Test-time Scaling}
\label{appendix:test-time-scaling}
For sequential scaling, the initial reasoning length of r1-qn-14b is around 1k tokens. When "Wait" is appended after the rationale to encourage the LLM to continue extending its reasoning, the accuracy of r1-qn-14b increases as the length approaches 2k, with a maximum improvement of 0.03. However, in the 2k–4k token range, the model's accuracy gradually declines due to a sharp drop in recall—sequential scaling induces excessive reflection, leading to a decrease in recall performance. Beyond 4k tokens, changes in accuracy begin to plateau, as the rationale cannot be extended indefinitely. When exceeding 4k tokens, the model often produces abnormal outputs, such as infinite repetition of previous content, which aligns with the findings of Muennighoff et al.~\cite{muennighoff2025s1}.

\section{Abnormal Output of LLMs}
\label{appendix:abnormaloutput}

\begin{figure}[H]
    \centering
    \hspace{-0.6cm}
    \includegraphics[width=0.45\textwidth]{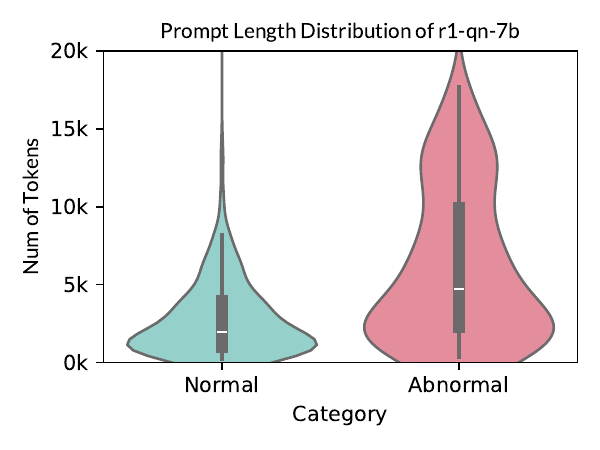}
    \caption{Prompt Length Distribution for Normal vs. Abnormal Outputs.}
    \label{fig:abnormal}
\end{figure}

The r1-qn-7b model exhibits an exceptionally high ``other'' rate of 32\% in \autoref{fig:pair-wise}. Through manual inspection, we identify two primary phenomena associated with the ``other'' outputs in r1-qn-7b:
\begin{itemize}[leftmargin=*, noitemsep, nolistsep]
    \item \textbf{Repetitive outputs}: Repeating a segment until reaching the output limit.
    \item \textbf{Non-compliance with instructions}: Failing to follow the expected format specified in the instruction.
\end{itemize}
Additionally, we observe that (1) as the model size increases, the occurrence of ``other'' cases decreases, and (2) longer inputs are more likely to result in anomalies in r1-qn-7b. To investigate this behavior further, we analyze the distribution of prompt lengths for cases where r1-qn-7b produces anomalous outputs compared to those where it does not. The average prompt lengths for normal and abnormal cases are 5,751 and 12,289, respectively, with median lengths of 3,925 and 9,466. The distribution is illustrated in \autoref{fig:abnormal}.

% ----------------------------------------
\section{Examples of Incorrect Reasoning Patterns}
\label{appendix:patterns}
\begin{figure}[H]
    \centering
    \includegraphics[width=0.475\textwidth]{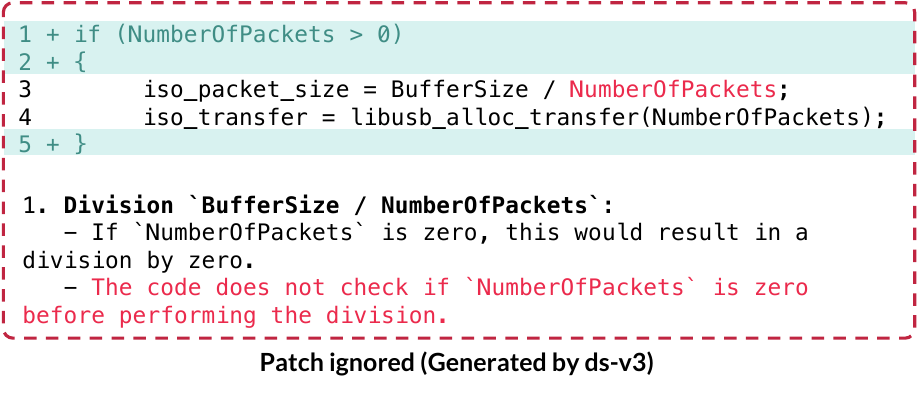}
    \caption{CVE-2022-39318, a division-by-zero vulnerability.}
    \label{fig:patchignoredexample}
\end{figure}

\noindent\textbf{Patch Ignored:}
\autoref{fig:patchignoredexample} illustrates one of the cases where ds-v3 fails due to "Patch Ignored". As shown, ds-v3 completely failed to recognize the presence of the patch \textsf{(NumberOfPackets > 0)} and output: "The code does not check if \textsf{NumberOfPackets} is zero before performing the division."

\begin{figure}[H]
    \centering
    \includegraphics[width=0.475\textwidth]{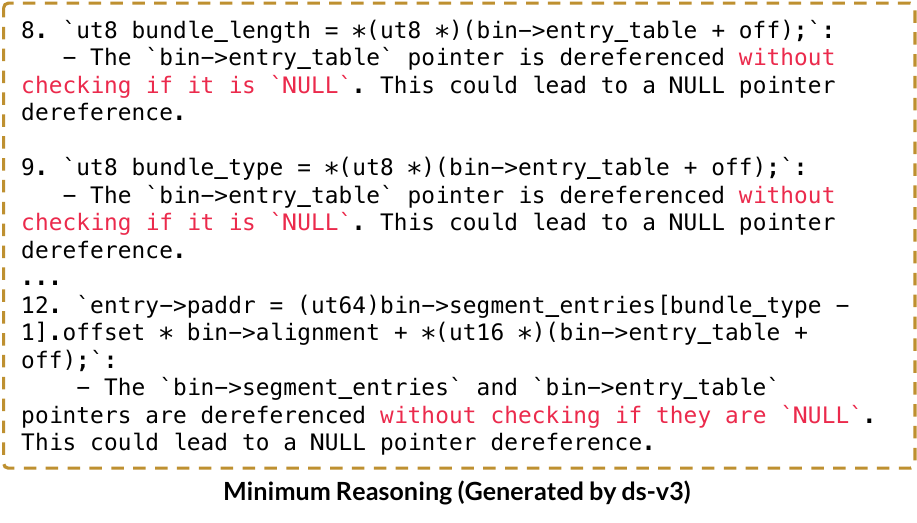}
    \caption{CVE-2022-1283, a null pointer dereference vulnerability.}
    \label{fig:minimumreasoning}
\end{figure}

\noindent\textbf{Minimum Reasoning:}
\autoref{fig:minimumreasoning} demonstrates an example where ds-v3 falters due to "Minimum Reasoning." As shown, ds-v3 provides almost no explanation for why it judged the case as a vulnerability, offering only a superficial analysis. In fact, a more detailed analysis would reveal that \textsf{bin->entry\_table} is not \textsf{NULL}.

\begin{figure}[H]
    \centering
    \includegraphics[width=0.475\textwidth]{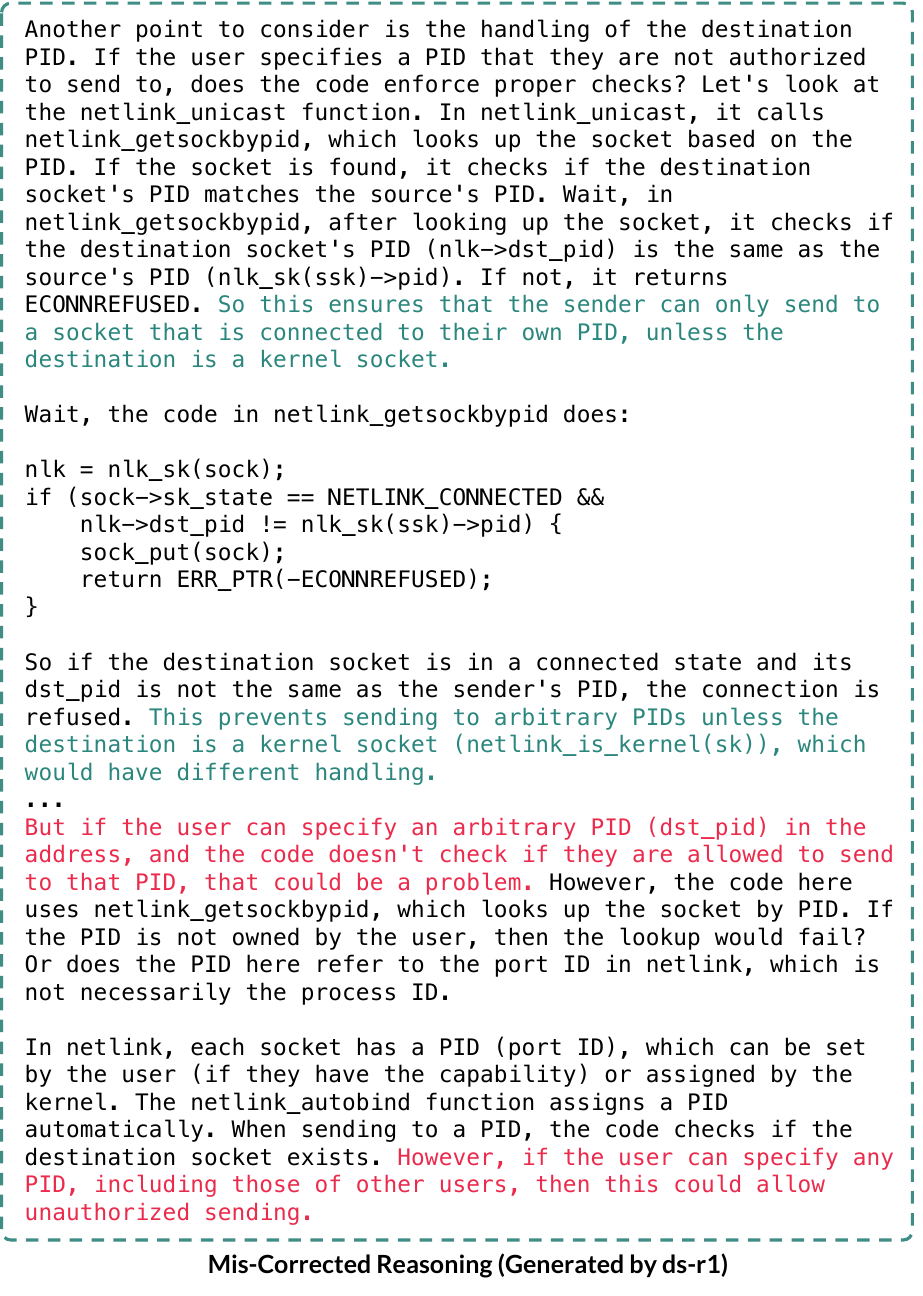}
    \caption{CVE-2012-6689, an improper access control vulnerability.}
    \label{fig:miscorrected}
\end{figure}

\noindent\textbf{Mis-Corrected:}
Figure \autoref{fig:miscorrected} presents an example where ds-r1 makes an error due to "Mis-Corrected Reasoning." Initially, ds-r1 correctly believes that the vulnerability is successfully prevented. However, during subsequent reasoning, ds-r1 reconsiders and incorrectly concludes that a vulnerability exists.

\end{document}